\documentclass[aps,pra,twocolumn,superscriptaddress,longbibliography]{revtex4-1}
\usepackage{graphicx}
\usepackage{latexsym}
\usepackage{amssymb}
\usepackage{amsmath}
\usepackage{amsfonts}
\usepackage{upgreek}
\usepackage{float}
\usepackage{bm}
\usepackage{multirow}
\usepackage{color}
\usepackage[T1]{fontenc}
\usepackage{hyperref}
\hypersetup{
colorlinks = true,
linkcolor = [rgb]{0.70,0.13,0.13},
citecolor = [rgb]{0.13,0.55,0.13},
urlcolor  = [rgb]{0.25, 0.41, 0.88}}
\newcommand{\ket}[1]{|#1\rangle}

\begin{document}

\title{Topological phases of spinless $p$-orbital fermions in zigzag optical lattices}
\author{Gaoyong Sun}
\thanks{Corresponding author: gysun@nuaa.edu.cn}
\affiliation{College of Science, Nanjing University of Aeronautics and Astronautics, Nanjing, 211106, China}
\author{Wen-Long You}
\thanks{Corresponding author: wlyou@nuaa.edu.cn}
\affiliation{College of Science, Nanjing University of Aeronautics and Astronautics, Nanjing, 211106, China}
\author{Tao Zhou}
\thanks{Corresponding author: tzhou@scnu.edu.cn}
\affiliation{Guangdong Provincial Key Laboratory of Quantum Engineering and Quantum Materials, GPETR Center for Quantum Precision Measurement, SPTE, and Frontier Research Institute for Physics, South China Normal University, Guangzhou 510006, China}

\begin{abstract}
Motivated by the experiment [St-Jean {\it et al}., Nature Photon. {\bf 11}, 651 (2017)] on topological phases with collective photon modes in a zigzag chain of polariton micropillars,
we study spinless $p$-orbital fermions with local interorbital hoppings and repulsive interactions between $p_x$ and $p_y$ bands in zigzag optical lattices.
We show that spinless $p$-band fermions in zigzag optical lattices can mimic the interacting Su-Schrieffer-Heeger model and the effective transverse field Ising model in the presence of local hoppings.
We analytically and numerically discuss the ground-state phases and quantum phase transitions of the model.
This work provides a simple scheme to simulate topological phases and the quench dynamics of many-body systems in optical lattices.

\end{abstract}

\maketitle

\section{Introduction}
Topological phases of matter are fascinating quantum states in modern condensed matter physics, which are characterized by some prominent features, such as string orders, robust edge states
beyond the Landau-Ginzburg symmetry-breaking theory \cite{levin2006detecting}.
The Su-Schrieffer-Heeger (SSH) model that exhibits topological nontrivial phases was originally proposed for fermionic particles with staggered hoppings in polyacetylene chains \cite{su1979solitons, su1980soliton}.
The SSH model is a simple but very important model in studying the topology of the single-particle band structure in solid-state physics.
Thanks to the rapid development of quantum simulations \cite{bloch2008many, lahaye2009physics, georgescu2014quantum}, the SSH model was recently realized in versatile platforms,
such as coupled semiconductor micropillars with the collective photon modes \cite{kruk2017edge, st2017lasing},
and optical lattices with ultracold atoms \cite{nakajima2016topological, lohse2016thouless, de2019observation, xie2019topological}.

A natural proposal to realize the SSH model in optical lattices is to create a double well superlattice with the same unit cell as the original SSH model \cite{nakajima2016topological, lohse2016thouless, de2019observation, xie2019topological}.
Interestingly, an orbital version of the SSH Hamiltonian was implemented by using polariton micropillars in a $p$-band zigzag chain in Ref.\cite{st2017lasing}, 
where the topological nontrivial phases and topological trivial phases were found to form in the orthogonal $p_x$ and $p_y$ subspaces.
However, the impact from the mixing of the $p_x$ and $p_y$ orbitals and the on-site interactions were not investigated in Ref.\cite{st2017lasing}, 
which we believe are important to engineer rich many-body physics in optical lattices. This is because:
(i) In cold atoms, it may be very difficult to prepare orthogonal $p_x$ and $p_y$ orbitals with perfect $90^{\circ}$ angles.
In fact, it would be interesting to investigate phase transitions of the $p$-band systems by introducing such deformations of the local lattice wells \cite{sun2012exploring} 
or local anharmonicity of the lattice potential \cite{pinheiro2013x,sowinski2013tunneling,saugmann2020magnetic} instead of considering only the orthogonal $p_x$ and $p_y$ subspaces. 
(ii) When placing the bosons or spinful fermions on $p$-band optical lattices, the pair hopping terms due to the Hund effect would cause a mixing of $p_x$ and $p_y$ orbitals of a given lattice well.
(iii) In the strong on-site interaction limit, a small mixing of orbitals may lead to a phase transition because the effective coupling strength from the second-order perturbation theory is small.

\begin{figure}
\includegraphics[width=7.3cm]{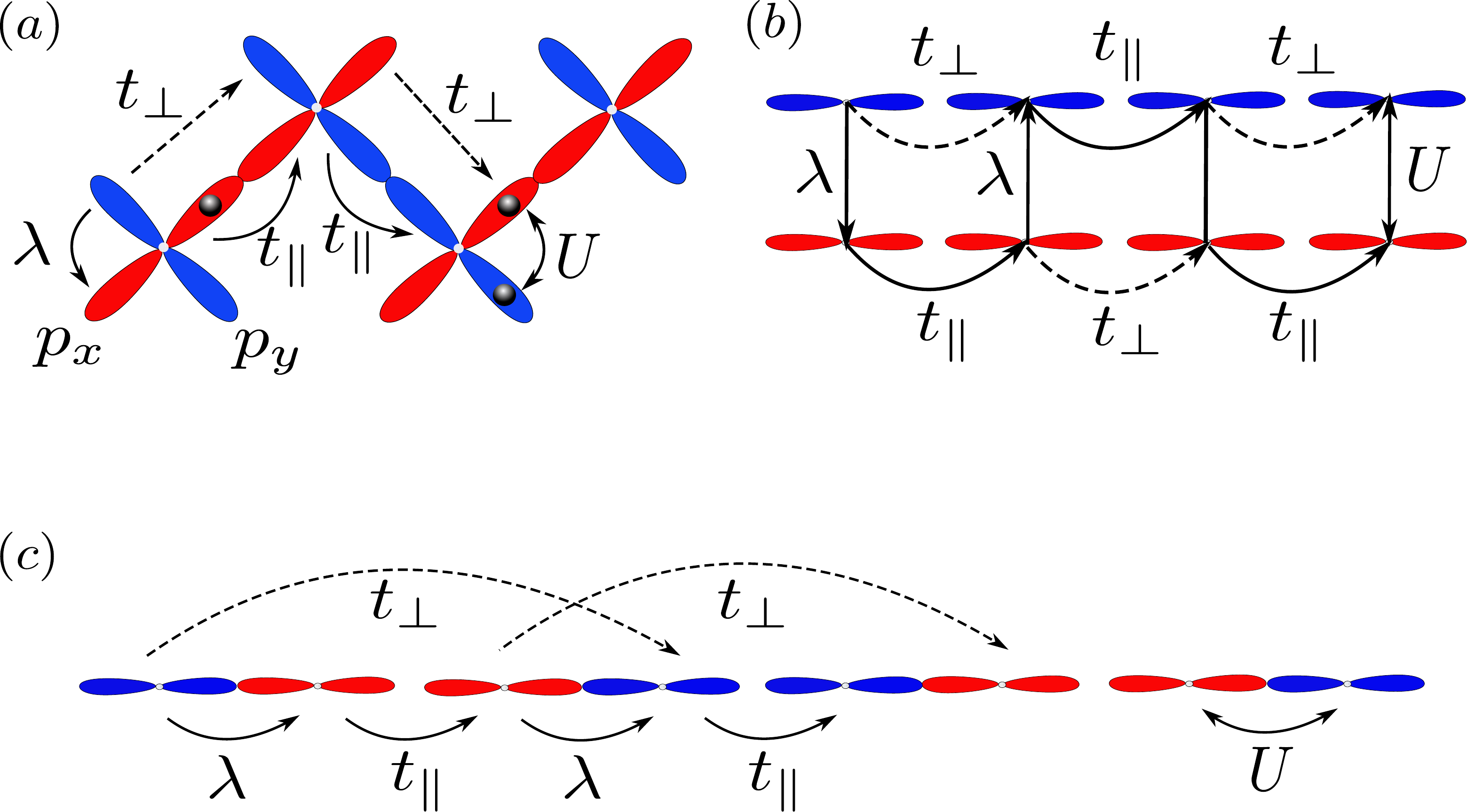}
\caption{ (Color online)
Geometry of the $p$-band model discussed in this work. (a) Zigzag lattice with degenerate $p_x$ and $p_y$ orbitals occupied by spinless fermions, 
where $t_{\parallel}$ and $t_{\perp}$ denote the longitudinal and transverse hopping between the same orbitals in nearest-neighboring lattice sites, 
$\lambda$ and $U$ refer to the local hopping and interaction between different orbitals in a given site.
(b) The equivalent ladder geometry of (a). 
(c) The representation for the Hamiltonian in Eq. (\ref{Ham}) in terms of spinless fermions on a SSH-like chain.}
\label{pbandfig}
\end{figure}

In this paper, we generalize the work of Ref.\cite{st2017lasing} that 
realizes the SSH model with polariton micropillars by considering spinless fermions loaded in a $p$-band zigzag optical lattice with the
on-site hopping (band mixing) and on-site interactions, which were discarded in Ref.\cite{st2017lasing}.
We show that the topological phases persist under such local deformations and the phase transition in the strong interacting limit at half-filling is described by the effective transverse field Ising model.
We note that the $p$-bands systems in optical lattices have been investigated experimentally \cite{isacsson2005multiflavor, muller2007state, wirth2011evidence, niu2018observation, slot2019p}
and theoretically  \cite{liu2006atomic, wu2007flat, wu2008orbital, zhao2008orbital, lu2009dispersive, wu2009unconventional, hauke2011orbital, kobayashi2012nontrivial, soltan2012quantum, li2012time, sun2012exploring, pinheiro2013x,sowinski2013tunneling,wu2012topological,li2013topological, sun2014ferromagnetic, you2014quantum, zhou2015spin, dutta2015non, li2016physics,xu2016pi, liu2018chiral, li2018rotation, jin2019dynamical, zhu2019interaction,saugmann2020magnetic}.

This paper is organized as follows.
In Sec.\ref{sec:model}, we introduce the $p$-band model with spinless fermions in zigzag optical lattices.
In Sec.\ref{sec:SPS}, we study the quantum phases without interactions by the single particle spectrum.
In Sec.\ref{sec:SCM}, we discuss the quantum phases and phase transitions in the presence of  strong interactions and derive the effective transverse field Ising model.
In Sec.\ref{sec:FPD}, we present the full phase diagram of the model.
Finally, in Sec.\ref{sec:Con}, we summarize this work.

\section{Model}
\label{sec:model}
We consider spinless fermions loaded in a zigzag optical lattice \cite{sun2012exploring, sun2014ferromagnetic, you2014quantum}
as shown in Fig.\ref{pbandfig}(a), where two degenerate $p_x$ and $p_y$ orbitals are active within the $x$-$y$ plane per lattice site due to a strong confinement along $z$-direction.
For spinless fermions, the $p$-orbital bands can be realized by Fermi statistics, where the lowest $s$-orbital band is fully filled (that can be removed afterward with laser pulses \cite{hauke2011orbital}).
The higher orbital bands, such as $d, f$ bands, are separated by large band gaps \cite{kobayashi2012nontrivial,li2016physics}.
Hence, only $p$-bands are active and inter-band effects are negligible. Consequently, the Hamiltonian of the system composed of $N$ lattice wells is given by \cite{liu2006atomic, zhao2008orbital},
\begin{align}
H = -{}& \sum_{i=1, l=1,2}^{N}  (t c^{\dagger}_{i, l} c_{i+1, l} + \lambda c^{\dagger}_{i, p_x} c_{i, p_y} + h.c.)  \nonumber \\
      +{}& \sum_{i=1}^{N}  U c^{\dagger}_{i, p_x} c_{i, p_x} c^{\dagger}_{i, p_y} c_{i, p_y},
\label{Ham}
\end{align}
with $t=-\frac{1}{2} t_{\parallel} [1+(-1)^{i+l}] + \frac{1}{2} t_{\perp} [1-(-1)^{i+l}]$, where $t_{\parallel}$ and $t_{\perp}$ are the longitudinal and transverse hopping amplitudes
along the same orbital $p_x$ (or $p_y$) between two nearest-neighbor lattice sites,
and $l=1,2$ indicates the $p_x, p_y$ orbital in a given lattice site. In Fig.\ref{pbandfig}(a), the longitudinal hopping $t_{\parallel}$ is much larger than the transverse hopping $t_{\perp}$
because the overlap integrals of hopping amplitudes are dependent on the orientation of orbitals \cite{liu2006atomic, zhao2008orbital}.
The local interorbital hopping $\lambda$ that leads to a mixing of the $p_x$ and $p_y$ orbitals can be tuned by a deformation of the lattice wells, such as by an additional weak tilted lattice \cite{sun2012exploring}.
Here $c^{\dagger}_{i, l}, c_{i, l} $ are the creation and annihilation operators at $l$th orbital of the $i$th site, and $U>0$ is the on-site repulsive interaction between $p_x$ and $p_y$ orbitals in a single given well. 
We note that the hopping coefficients, the interactions and the band gaps are dependent on the lattice depth of the optical wells \cite{bloch2008many, lahaye2009physics,wirth2011evidence,liu2006atomic}. 
The interactions of fermions can be tuned independently by the $p$-wave Feshbach resonance (not the $s$-wave Feshbach resonance for bosons) due to the Pauli exclusion principle \cite{hauke2011orbital}.
It is easy to find that the $p$-band model in zigzag lattices is equivalent to a spinless fermionic model on a two-leg ladder [cf. Fig.\ref{pbandfig}(b)] or
a one-dimensional SSH chain [cf. Fig.\ref{pbandfig}(c)], which we will discuss in more detail below.

\begin{figure} [t!]
\includegraphics[width=7.0cm]{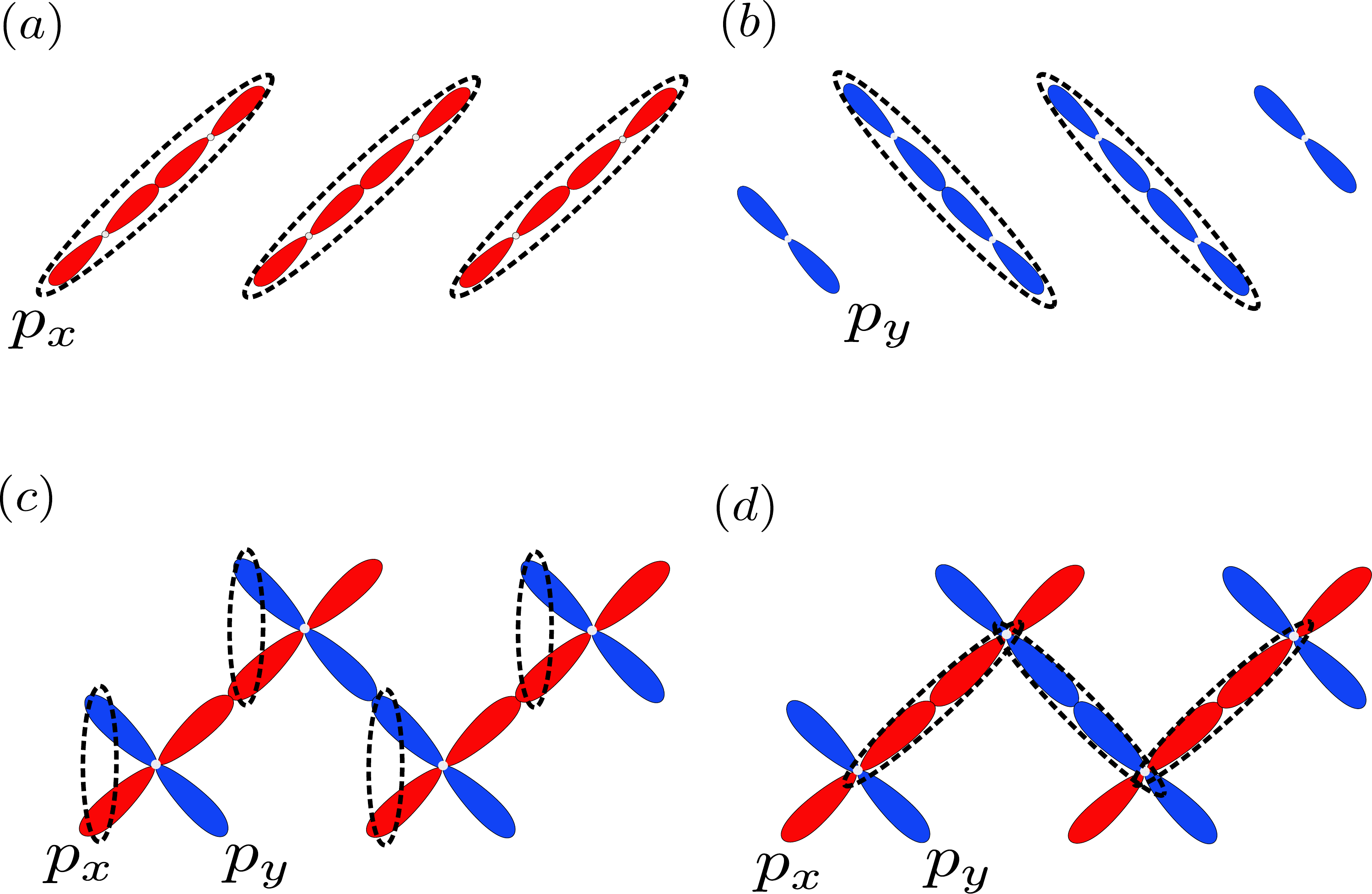}
\caption{(Color online)
Quantum phases of non-interacting Hamiltonian $U=0$. (a) Trivial orbital phase in the subspace of $p_x$ orbitals with $\lambda=0$ and $t_{\perp} < t_{\parallel}$,
(b) Topological orbital phase in the subspace of $p_y$ orbitals with $\lambda=0$ and $t_{\perp} < t_{\parallel}$,
(c) Trivial orbital phase with the configuration of the superposition of $p_x$ and $p_y$ orbitals at all local lattice wells with $t_{\perp} = 0$ and $\lambda > t_{\parallel}$,
(d) Topological orbital phase with the configuration of the superposition of same orbitals between two nearest-neighbor lattice wells with $t_{\perp} = 0$ and $\lambda < t_{\parallel}$.}
\label{pbandfig2}
\end{figure}

\section{Topological phases in the non-interacting SSH-like model}
\label{sec:SPS}
Let us first consider the noninteracting case ($U=0$) of Hamiltonian in Eq.(\ref{Ham}).  
In the absence of interorbital hopping ($\lambda=0$), 
the $p_x$ orbitals and the $p_y$ orbitals are decoupled into two independent chains (subspaces) with staggered $t_{\parallel}$ and $t_{\perp}$ hopping as shown in Fig.\ref{pbandfig}(b).
In the chains with open boundary conditions, considering the longitudinal hopping $t_{\parallel}$ is typically much larger than the transverse hopping $t_{\perp}$ due to the orientation of orbitals \cite{liu2006atomic, zhao2008orbital},
the $p_x$ subspace consequently exhibits a dimerization 
on the $(2i-1, 2i)$ bonds without edge states as shown in Fig.\ref{pbandfig2}(a),
while the $p_y$ subspace forms a dimerization 
on the $(2i, 2i+1)$ bonds with topological edge states as demonstrated in Fig.\ref{pbandfig2}(b).
The odd-bond dimerizations in $p_x$ subspaces correspond to the topological trivial phase, while the even-bond dimerization in $p_y$ subspaces exhibit the topological nontrivial phase of the SSH model 
that was experimentally investigated with polariton micropillars in Ref.\cite{st2017lasing}.

Next, we consider the effect of the orbital deformation that introduces the on-site interorbital hopping ($\lambda \neq 0$) between the $p_x$ and $p_y$ orbitals of the local lattice wells, which
was neglected in Ref.\cite{st2017lasing}. In this context,
we arrive at the following noninteracting Hamiltonian with periodicity two [see Fig.\ref{pbandfig}(a)] by considering only the leading terms,
\begin{align}
H^{\prime} = -& \sum_{i=1,l=1,2}^{N}  (  t^{\prime} c^{\dagger}_{i, l} c_{i+1, l} + \lambda c^{\dagger}_{i, p_x} c_{i, p_y} + h.c.),
\label{HamSSH}
\end{align}
where $t^{\prime}=-\frac{1}{2} t_{\parallel} [1+(-1)^{i+l}]$, and we have discarded the transverse hopping term $t_{\perp}$ because $t_{\perp} \ll t_{\parallel}$.

\begin{figure} [t!]
\includegraphics[width=8.8cm]{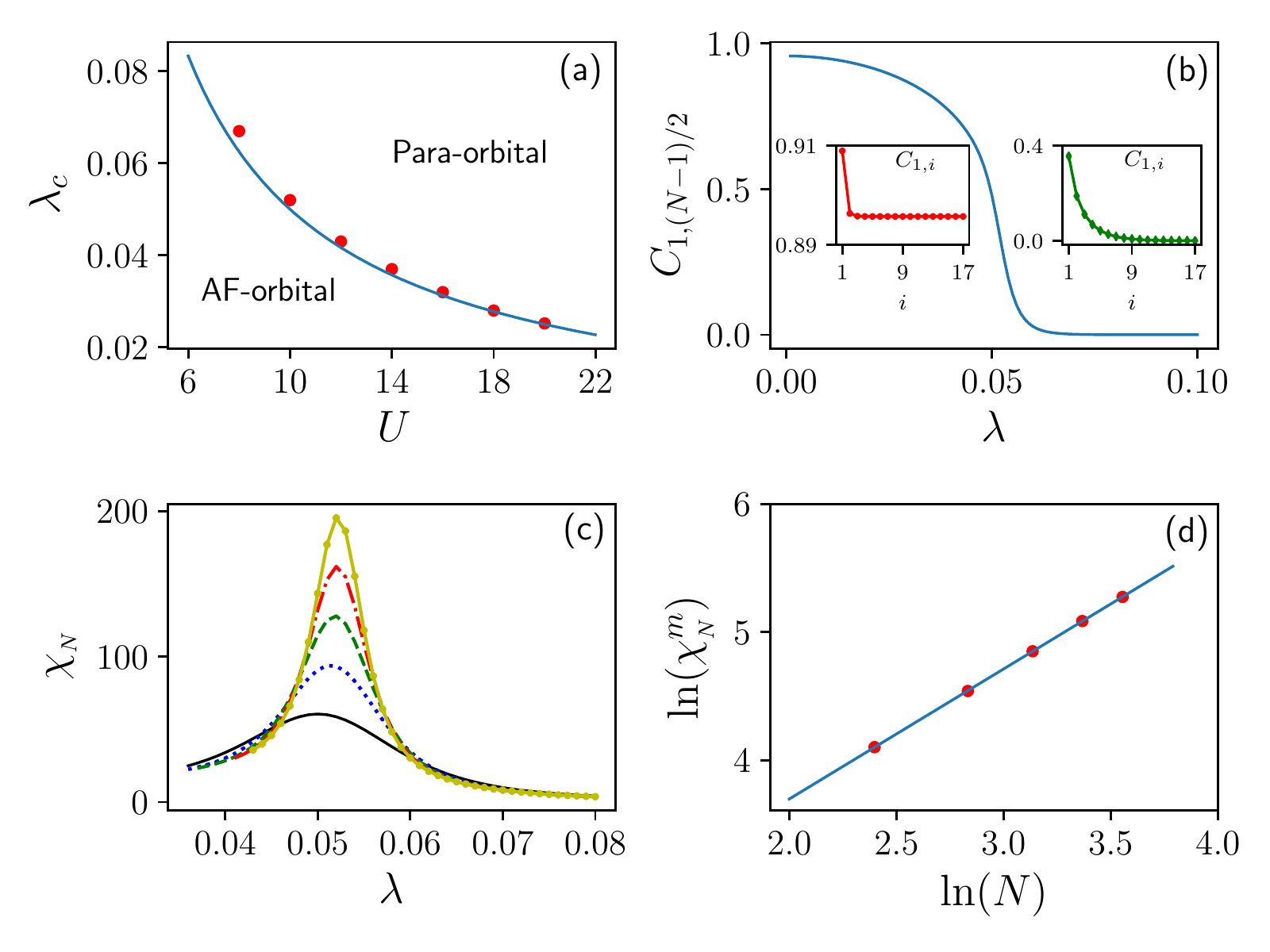}
\caption{(Color online)
Effective antiferro-orbital Ising model in strong interaction limit $U \gg t_{\parallel}$ with $t_{\perp} = 0$ and $t_{\parallel}=1$.
(a) Critical points $\lambda_c$ with respect to the interactions $U$ for $N=35$ wells ($N^{\prime}=2N=70$ orbitals in Fig.\ref{pbandfig}(c)) at half-filling.
The red filled circles are the numerical data obtained by the peak of fidelity susceptibility as shown in (c) from the original model Eq.({\ref{Ham}}), 
and the blue solid line denotes the exact analytical results $\lambda_c=t_{\parallel}^{2}/2U$ from the effective model Eq.({\ref{HamEff}}).
(b) The correlation function $C_{1,(N-1)/2}$ as a function of $\lambda$ with $N=35$ wells at $U=10$. Inset figures show the correlation function  $C_{1,i}$ with respect to the site $i$ at $\lambda=0.025$ (left) and $\lambda=0.075$ (right).
(c) The fidelity susceptibility per orbital with $N=11$ (solid line), $N=17$ (dotted line), $N=23$ (dashed line), $N=29$ (dash-dotted line), and $N=35$ (filled-circle line) wells at $U=10$.
(d) The finite-size scaling of the maximal values of the fidelity susceptibility in (c) with $N=11,17,23,29,35$ at $U=10$.}
\label{largeUfig}
\end{figure}

The Bogoliubov-de Gennes (BdG) Hamiltonian of Eq.(\ref{HamSSH}) under periodic boundary conditions can be easily derived as,
\begin{align}
H^{\prime} =
\begin{pmatrix}
a_k^{\dagger}, b_k^{\dagger}, c_k^{\dagger}, d_k^{\dagger} \\
\end{pmatrix}
\mathcal{H}(k)
\begin{pmatrix}
a_k \\
b_k \\
c_k \\
d_k
\end{pmatrix},
\label{HamBdG}
\end{align}
with,
\begin{align}
\mathcal{H}(k) =
\begin{pmatrix}
0               & -\lambda  & 0             & t_{\parallel} e^{-ik} \\
-\lambda   & 0             & t_{\parallel} & 0 \\
0               & t_{\parallel} & 0             & -\lambda \\
t_{\parallel} e^{ik} & 0             & -\lambda  & 0
\end{pmatrix},
\label{HamBdG}
\end{align}
by using the Nambu basis $\psi_{k}^{T}= (a_k, b_k, c_k, d_k)$. Here $a_k$, $b_k$, $c_k$, $d_k$ are the annihilation operators in the momentum space of $c_{2i-1, p_y}$, $c_{2i-1, p_x}$, $c_{2i, p_x}$, $c_{2i, p_y}$.
Diagonalizing the Hamiltonian (\ref{HamBdG}), we obtain the energy spectrum of the bulk states:
\begin{align}
\label{Ek1}
E(k) = \pm \sqrt{ \lambda^{2} + t_{\parallel}^{2} \pm  2 \lambda t_{\parallel} \cos{(k/2)}},
\end{align}
with $k = \frac{2\pi}{N/2} j $ and $j=1,2, \dotsb, N/2$. 
The Eq.(\ref{Ek1}) can be re-written as two bands energy spectrum of SSH model,
\begin{align}
E(k^{\prime}) = \pm \sqrt{ \lambda^{2} + t_{\parallel}^{2} +  2 \lambda t_{\parallel} \cos{(k^{\prime})}},
\end{align}
with $k^{\prime}$=$\frac{k}{2} \bigcup \frac{k}{2}+\pi$ because $-\cos(k/2) = \cos(k/2+\pi)$.
Obviously, if the degenerate $p_x$ and $p_y$ orbitals are 
regarded as two sublattice sites in each unit cell, one can easily arrive at the standard SSH model [see Fig.\ref{pbandfig}(c)].
Consequently, when $\lambda > t_{\parallel}$, a dimerized state 
is formed between the $p_{x}$ and $p_y$ sublattices in each single well as shown in Fig.\ref{pbandfig2}(c), which
corresponds to the topological trivial phase of SSH model. 
When $\lambda < t_{\parallel}$, the system exhibits
the dimerization between the $p_{x}$ orbitals on the odd bonds and the $p_y$ orbitals on the even bonds, respectively,
leaving the first and last $p_y$ orbitals as two edge states in the case of open boundary conditions [see Fig.\ref{pbandfig2}(d)].
To this end, there is a quantum phase transition between the topological trivial phase and the topological nontrivial phase at $\lambda=t_{\parallel}$ that can be reached by tuning the deformation induced interorbital hopping $\lambda$.
We note that when the transverse hopping terms $t_{\perp}$ are finite, the model becomes a SSH model with the third-neighbor hopping \cite{lee2016anomalous, yao2018edge}.
However, it would not qualitatively change the underlying physics for $t_{\perp}=0$
because the hopping strength $t_{\perp}$ is much smaller than $t_{\parallel}$ due to the orientation of orbitals \cite{liu2006atomic, zhao2008orbital}.

\begin{figure} [t!]
\includegraphics[width=8.8cm]{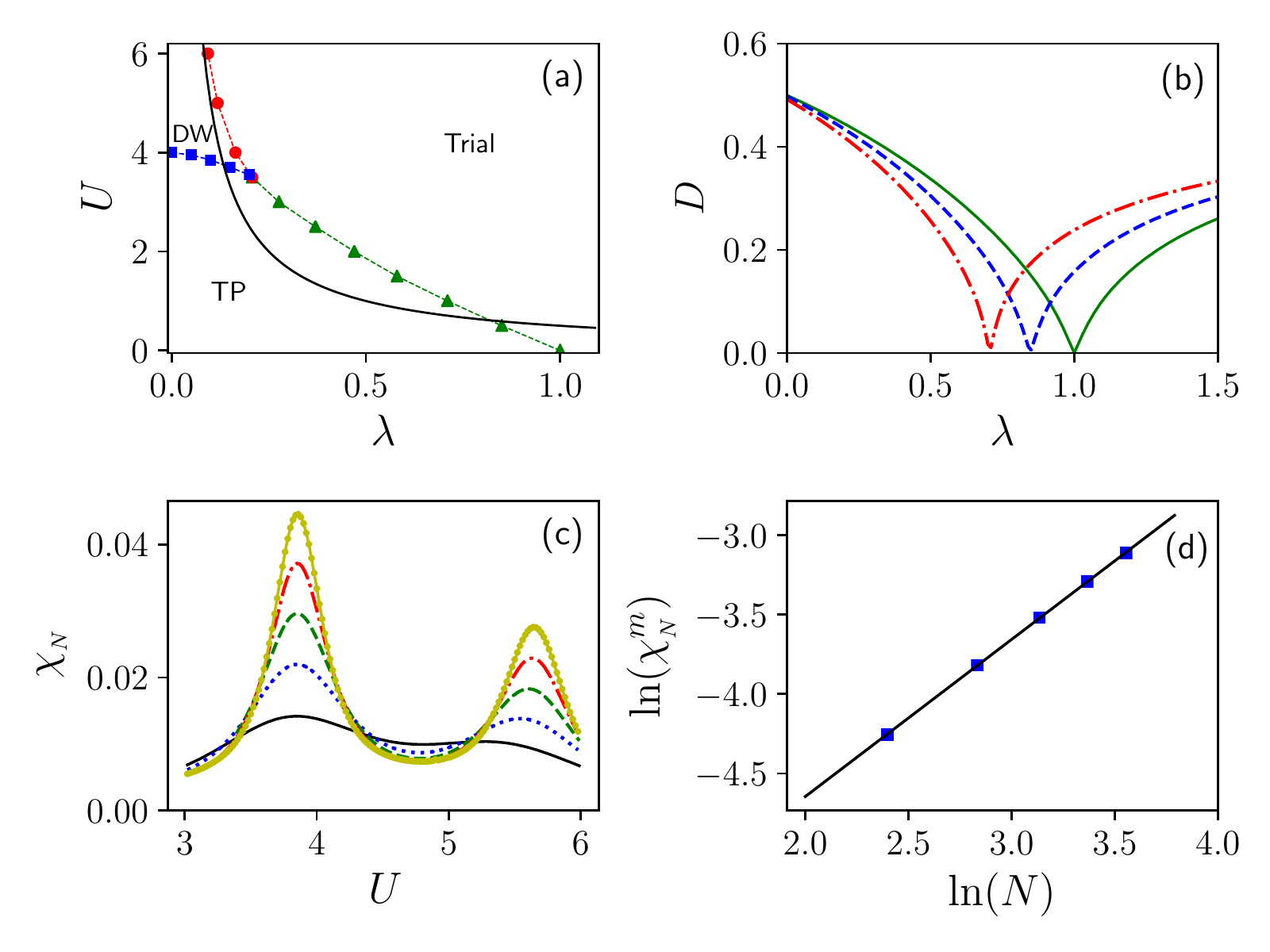}
\caption{(Color online)
(a) Full phase diagram of the model Eq.(\ref{Ham}) with respect to the interaction $U$ and the local interorbital hopping $\lambda$ 
with $t_{\perp} = 0$ and $t_{\parallel}=1$ for $N=35$ wells ($N^{\prime}=2N=70$ orbitals in Fig.\ref{pbandfig}(c)) at half-filling.
The filled symbols are the numerical data obtained from the original model Eq.({\ref{Ham}}) by the DMRG, 
and the black solid line denotes the exact analytical results $\lambda_c=t_{\parallel}^{2}/2U$ from the effective model Eq.({\ref{HamEff}}).
(b) The dimerized order $D$ as a function of $\lambda$ with $N=35$ wells at $U=0$ (green solid line), $U=0.5$ (blue dashed line) and $U=1$ (red dash-dotted line).
(c) The fidelity susceptibility per orbital with $N=11$ (solid line), $N=17$ (dotted line), $N=23$ (dashed line), $N=29$ (dash-dotted line), and $N=35$ (filled-circle line) wells at $\lambda=0.1$.
(d) The finite-size scaling of the maximal values (for the left peak around $U=4$) of the fidelity susceptibility in (c) with $N=11,17,23,29,35$ at $\lambda=0.1$. The critical exponent of the correlation length $\nu \approx 0.99$.
The phase transition denoted by the right peak of the fidelity susceptibility in (c) was discussed in Fig.(\ref{largeUfig}) with $\nu \approx 1.01$.}
\label{fullfig}
\end{figure}

\section {Effective strong-coupling model}
\label{sec:SCM}
In the following, we will study the ground-state properties and the associated quantum phase transitions of Hamiltonian in Eq.(\ref{Ham}) with on-site repulsive interaction $U \neq 0$. 
For simplicity but without loss of generality, we still overlook the transverse hopping $t_{\perp}$ terms in the following discussions.
The model is then the usual SSH model with the nearest-neighbor interaction $U$ between the $p_x$ and $p_y$ orbitals within a unit cell. 
To understand the nature of the quantum phases and the phase transitions of the $p$-band model in Eq.(\ref{Ham}),
we derive an effective antiferro-orbital (AF-orbital) Ising model in the strongly interaction limit with $U \gg t_{\parallel}$, 
by the second-order perturbation theory at half-filling \cite{liu2006atomic, zhao2008orbital, sun2012exploring, sun2014ferromagnetic}:
 \begin{align}
 H_{\text{eff}} = {}& \sum_{i=1}^{N} (J S_{i}^{z} S_{i+1}^{z} - 2 \lambda S_{i}^{x}),
 \label{HamEff}
 \end{align}
where $J=2t_{\parallel}^2/U$, $S_{i}^{\dagger}=c_{i, p_x}^{\dagger}c_{i, p_y}$ and $S_{i}^{z}=(c_{i, p_x}^{\dagger}c_{i, p_x} - c_{i, p_y}^{\dagger}c_{i, p_y})/2$. 
Hence, for $\lambda > t_{\parallel}^{2}/2U$, it is a 
para-orbital phase, while for $\lambda < t_{\parallel}^{2}/2U$, it is an antiferro-orbital Ising phase $(p_x, p_y, p_x, p_y, \dotsb)$,
in which one particle is located in the $p_x$ orbital of $i$th well and the other dwells on the $p_y$ orbital of the nearest neighbor $(i+1)$th well.
We note that in contrast to SU(2) symmetric Heisenberg interactions in spin models, the orbital exchange Hamiltonian evokes Ising-type interactions without quantum fluctuations, similar to the systems with $t_{2g}$
orbital degeneracy \cite{PhysRevB.78.214423, Daghofer2008absence}. 
Especially, the interorbital hopping $\lambda$ herein is responsible for substantial quantum fluctuations and plays a role of an external transverse field, 
which is hardly experimentally controlled in the orbital-only models of Mott insulators \cite{PhysRevB.78.214423, Daghofer2008absence}.
To verify our theoretical analysis, we compute the correlation function,
\begin{align} 
C_{1,i}=|\langle S_{1}^{z}S_{i}^{z}\rangle|, 
\end{align}
and the fidelity susceptibility per orbital \cite{PhysRevE.76.022101,gu2010fidelity,sun2017fidelity, zhu2018fidelity},
\begin{align}
\chi_{L}(\lambda)=\frac{1}{2N} \lim_{\delta \lambda \rightarrow 0} \frac{-2\ln F(\lambda,\lambda+\delta \lambda)}{(\delta \lambda)^2},
\label{eqFS}
\end{align}
Where, the correlation function $C_{1,i}$ tends to the square $C_{1,i} \rightarrow m_{z}^{2}$ of local order parameter $m_z=\frac{1}{N} \sum_{i}(-1)^{i} \langle S_{i}^{z} \rangle$ for large distance $i$,
and the fidelity $F(\lambda,\lambda+\delta \lambda)$=$\vert \langle \psi_0(\lambda)\vert  \psi_0(\lambda+\delta \lambda)\rangle\vert$ evaluates the overlap of two infinitesimally close states.

\begin{figure} [t!]
\includegraphics[width=8.8cm]{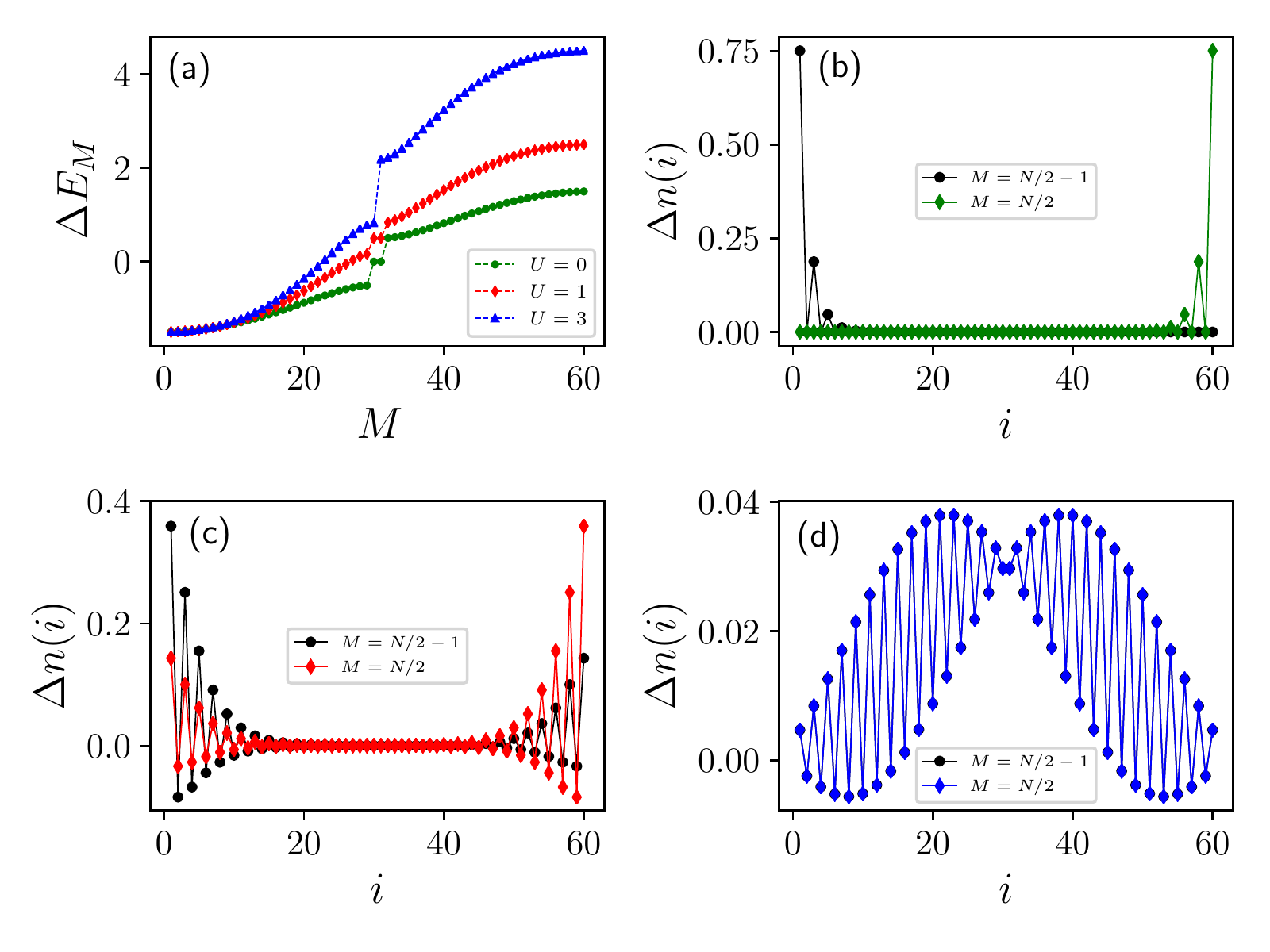}
\caption{(Color online)
(a) The quasi-particle energy spectrum $\Delta E_{M}$ with respect to the density $M$ of the model Eq.(\ref{Ham}) for $U=0$, $U=1$ and $U=3$ with $t_{\perp} = 0$, $t_{\parallel}=1$ and $\lambda=0.5$ 
for $N=30$ wells ($N^{\prime}=2N=60$ orbitals in Fig.\ref{pbandfig}(c)) with open boundary conditions.
The density distribution $\Delta n(i)$ of the quasi-particle of the two in-gap modes computed with the density $M=N/2-1$ and $M=N/2$ for (b) $U=0$, (c) $U=1$ and (d) $U=3$.}
\label{BBCfig}
\end{figure}

The numerical results are presented in Fig.\ref{largeUfig} by performing the density matrix renormalization group (DMRG) with periodic boundary conditions up to $N=35$ wells 
[equivalent to $N^{\prime}=70$ orbitals of Fig.\ref{pbandfig}(c)]. 
As is shown in Fig.\ref{largeUfig}(b), one can see clearly that a quantum phase transition occurs between
the antiferro-orbital Ising phase ($\lambda < \lambda_c$) and the para-orbital phase ($\lambda > \lambda_c$).
The critical point can be detected by the peak of the fidelity susceptibility as shown in Fig.\ref{largeUfig}(c).
The dependence of the critical values $\lambda_c$ on $U$ obtained from the original model Eq.({\ref{Ham}}) is presented in Fig.\ref{largeUfig}(a), which agrees well with
the analytical results $\lambda_c=t_{\parallel}^{2}/2U$ from the effective Ising model. 
Regarding the finite-size scaling of the peak of the fidelity susceptibility for a continuous phase transitions in one-dimensional system \cite{gu2010fidelity,PhysRevB.101.094410,PhysRevB.102.024425, sun2017fidelity, zhu2018fidelity},
\begin{eqnarray}
\chi_{N}^{m} \propto N^{2/\nu-1},
\end{eqnarray}
we obtain the critical exponent of the correlation length $\nu \approx 1.01$ consistent with Ising transition $\nu=1$ from 
maximal values of the fidelity susceptibility as shown in Fig.\ref{largeUfig}(d).
The effective Ising model in Eq.({\ref{HamEff}}) we derived allows for the simulations of the Ising transition 
or dynamical quantum phase transitions \cite{heyl2013dynamical, heyl2018dynamical} of the Ising model with strong interacting spinless fermions in zigzag lattices.

\section {Full phase diagram}
\label{sec:FPD}
In this section, we demonstrate the full phase diagram, which is presented in Fig.\ref{fullfig}(a) by the DMRG method up to $N=35$ sites with periodic boundary conditions.
In addition to the topological nontrivial phase (TP) and the trivial phase, a density wave phase (DW) (or Aoki phase in the Gross-Neveu model, antiferro-orbital phase in effective Ising model Eq.({\ref{HamEff}})) 
appears \cite{junemann2017exploring,bermudez2018gross, kuno2019phase, kuno2020interaction}
owing to the presence of the local interaction $U$.  The phase transition between the topological nontrivial phase and trivial phase can be characterized by a dimerized order,
\begin{align}
D =  \frac{1}{N} \sum_{i=1}^{N/2} {}& ( |c^{\dagger}_{2i-1, p_x} c_{2i-1, p_y} - c^{\dagger}_{2i-1, p_x} c_{2i, p_x}| \nonumber \\
    + {}& |c^{\dagger}_{2i, p_x} c_{2i, p_y} - c^{\dagger}_{2i, p_y} c_{2i+1, p_y}| ),
\end{align}
which becomes zero at the critical points as shown in Fig.\ref{fullfig}(b). The phase transitions for the density wave phase to the topological nontrivial phase and trivial phase are described by the fidelity susceptibility
that shows a double-hump structure [cf. Fig.\ref{fullfig}(c)]. Both of the phase transitions belong to the Ising universal class \cite{bermudez2018gross, kuno2019phase}
with the critical exponent of the correlation length $\nu  \approx 1$ as demonstrated in Fig.\ref{largeUfig}(d) and Fig.\ref{fullfig}(d).
However, we note that the effective Ising mode we derived is only valid for large $U$ [cf. Fig.\ref{fullfig}(a)].

To illustrate the nontrivial topological properties of the system with interactions, we introduce the quasiparticle energy spectrum $\Delta E_{M}$ as \cite{zhu2013topological,deng2014topological,sun2016topological,zhang2020skin},
\begin{eqnarray}
\Delta E_{M} = E_{M+1} - E_{M},
\end{eqnarray}
with the $E_{M}$ is the many-body ground-state energy of $M$ particles.
The energy spectra $\Delta E_{M}$ for $L=30$ sites with open boundary conditions are shown in Fig.\ref{BBCfig}(a), in which we find two in-gap modes
 in the topological nontrivial phase regime at $U=0$ and $U=1$ in contrast to the trivial phase regime at $U=3$. 
 The appearance of the in-gap modes is usually associated with the edge states localized at the boundaries,
 which can be confirmed by computing the density distribution $\Delta n(i)$ \cite{zhu2013topological,deng2014topological,sun2016topological,zhang2020skin},
 \begin{eqnarray}
\Delta n(i) = \left\langle \psi^{g}_{M+1} \middle| n_i  \middle| \psi^{g}_{M+1} \right\rangle - \left\langle \psi^{g}_{M} \middle| n_i  \middle| \psi^{g}_{M} \right\rangle,
\end{eqnarray}
 of the quasi-particle of the two in-gap modes by using the density $M=N/2-1$ and $M=N/2$, where the $n_i$ is the particle number operators and $\ket{\psi^{g}_{M}}$ is the many-body ground-state wave-function of $M$ particles.
 The density distribution $\Delta n(i)$ for $U=0$, $U=1$ and $U=3$ are presented in Fig.\ref{BBCfig}(b), Fig.\ref{BBCfig}(c), Fig.\ref{BBCfig}(d), where the edge states indeed appear in such in-gap modes.
 We note that topological phases and phase transitions can be studied by measuring the occupancy of bulk sites and edge sites in experiments \cite{de2019observation}.

\section {Conclusion}
\label{sec:Con}
In summary, we have shown that spinless fermions loaded in a $p$-band zigzag optical lattice can engineer the interacting SSH model, which shows a topological phase transition from the trivial phase to the topological nontrivial phase, 
where the edge states appear in open boundary conditions. In the strong interaction limit, the transverse field Ising model can be mimicked owing to the on-site band mixing and repulsion. 
We show the spinless fermions in $p$-band zigzag lattice can 
host rich quantum phases and the associated phase transitions due to the interplay between the lattice geometry, the deformation of the lattice wells and the interactions.

In addition, when the dipolar particles are loaded into the lattices, one may simulate the long-range interacting SSH and long-range Ising models \cite{lahaye2009physics}.
Consequently, our proposal opens a simple way to study quantum phase transitions and the quench dynamics, such as dynamical quantum phase transitions with broken symmetries \cite{sun2020dynamical}
of many-body systems. We note that it may also be possible to simulate a non-Hermitian 
SSH model or a non-Hermitian Ising model if the gain and loss are introduced into the systems \cite{ashida2020non, pickup2020synthetic}.
Moreover, we would expect a similar phase diagram for hard-core bosons in the zigzag optical lattices. 
However, it would be more interesting to understand the Hund effects and find additional effects with soft-core bosons in the future.

Finally, we would like to emphasize that the orbital symmetry in $p$-band zigzag lattice leads to $z$-component Ising interactions along any direction in the $xy$ plane in the regime of large $U$, i.e., for $U \gg t_{\parallel}$, 
in stark contrast to the ferromagnetic Kitaev interactions with nonequivalent components of Ising superexchange along different axes \cite{PhysRevLett.102.017205,PhysRevB.89.104425,PhysRevB.82.104416}. 
We note that the hole propagation described by the $tJ_{z}$ model in antiferro-orbital and para-orbital background may lead to a nontrivial many-body effect \cite{PhysRevB.78.214423, emery1990phase}.

\begin{acknowledgments}
G. S. is appreciative of support from the NSFC under the Grant Nos. 11704186 and 11874220.  W.-L. You acknowledges support by the startup fund of Nanjing University of Aeronautics and Astronautics under Grant No. 1008-YAH20006. 
T. Z. is supported by the NSFC under the Grant No. 12074130 and Science and Technology Program of Guangzhou under the Grant No. 2019050001.
Numerical simulations were performed on the clusters at Nanjing University of Aeronautics and Astronautics, and National Supercomputing Center in Shenzhen. 
\end{acknowledgments}

\bibliographystyle{apsrev4-1}
\bibliography{ref}

\begin{thebibliography}{70}%
\makeatletter
\providecommand \@ifxundefined [1]{%
 \@ifx{#1\undefined}
}%
\providecommand \@ifnum [1]{%
 \ifnum #1\expandafter \@firstoftwo
 \else \expandafter \@secondoftwo
 \fi
}%
\providecommand \@ifx [1]{%
 \ifx #1\expandafter \@firstoftwo
 \else \expandafter \@secondoftwo
 \fi
}%
\providecommand \natexlab [1]{#1}%
\providecommand \enquote  [1]{``#1''}%
\providecommand \bibnamefont  [1]{#1}%
\providecommand \bibfnamefont [1]{#1}%
\providecommand \citenamefont [1]{#1}%
\providecommand \href@noop [0]{\@secondoftwo}%
\providecommand \href [0]{\begingroup \@sanitize@url \@href}%
\providecommand \@href[1]{\@@startlink{#1}\@@href}%
\providecommand \@@href[1]{\endgroup#1\@@endlink}%
\providecommand \@sanitize@url [0]{\catcode `\\12\catcode `\$12\catcode
  `\&12\catcode `\#12\catcode `\^12\catcode `\_12\catcode `\%12\relax}%
\providecommand \@@startlink[1]{}%
\providecommand \@@endlink[0]{}%
\providecommand \url  [0]{\begingroup\@sanitize@url \@url }%
\providecommand \@url [1]{\endgroup\@href {#1}{\urlprefix }}%
\providecommand \urlprefix  [0]{URL }%
\providecommand \Eprint [0]{\href }%
\providecommand \doibase [0]{http://dx.doi.org/}%
\providecommand \selectlanguage [0]{\@gobble}%
\providecommand \bibinfo  [0]{\@secondoftwo}%
\providecommand \bibfield  [0]{\@secondoftwo}%
\providecommand \translation [1]{[#1]}%
\providecommand \BibitemOpen [0]{}%
\providecommand \bibitemStop [0]{}%
\providecommand \bibitemNoStop [0]{.\EOS\space}%
\providecommand \EOS [0]{\spacefactor3000\relax}%
\providecommand \BibitemShut  [1]{\csname bibitem#1\endcsname}%
\let\auto@bib@innerbib\@empty
\bibitem [{\citenamefont {Levin}\ and\ \citenamefont
  {Wen}(2006)}]{levin2006detecting}%
  \BibitemOpen
  \bibfield  {author} {\bibinfo {author} {\bibfnamefont {M.}~\bibnamefont
  {Levin}}\ and\ \bibinfo {author} {\bibfnamefont {X.-G.}\ \bibnamefont
  {Wen}},\ }\href {\doibase 10.1103/PhysRevLett.96.110405} {\bibfield
  {journal} {\bibinfo  {journal} {Physical Review Letters}\ }\textbf {\bibinfo
  {volume} {96}},\ \bibinfo {pages} {110405} (\bibinfo {year}
  {2006})}\BibitemShut {NoStop}%
\bibitem [{\citenamefont {Su}\ \emph {et~al.}(1979)\citenamefont {Su},
  \citenamefont {Schrieffer},\ and\ \citenamefont {Heeger}}]{su1979solitons}%
  \BibitemOpen
  \bibfield  {author} {\bibinfo {author} {\bibfnamefont {W.~P.}\ \bibnamefont
  {Su}}, \bibinfo {author} {\bibfnamefont {J.~R.}\ \bibnamefont {Schrieffer}},
  \ and\ \bibinfo {author} {\bibfnamefont {A.~J.}\ \bibnamefont {Heeger}},\
  }\href {\doibase 10.1103/PhysRevLett.42.1698} {\bibfield  {journal} {\bibinfo
   {journal} {Physical Review Letters}\ }\textbf {\bibinfo {volume} {42}},\
  \bibinfo {pages} {1698} (\bibinfo {year} {1979})}\BibitemShut {NoStop}%
\bibitem [{\citenamefont {Su}\ \emph {et~al.}(1980)\citenamefont {Su},
  \citenamefont {Schrieffer},\ and\ \citenamefont {Heeger}}]{su1980soliton}%
  \BibitemOpen
  \bibfield  {author} {\bibinfo {author} {\bibfnamefont {W.~P.}\ \bibnamefont
  {Su}}, \bibinfo {author} {\bibfnamefont {J.~R.}\ \bibnamefont {Schrieffer}},
  \ and\ \bibinfo {author} {\bibfnamefont {A.~J.}\ \bibnamefont {Heeger}},\
  }\href {\doibase 10.1103/PhysRevB.22.2099} {\bibfield  {journal} {\bibinfo
  {journal} {Physical Review B}\ }\textbf {\bibinfo {volume} {22}},\ \bibinfo
  {pages} {2099} (\bibinfo {year} {1980})}\BibitemShut {NoStop}%
\bibitem [{\citenamefont {Bloch}\ \emph {et~al.}(2008)\citenamefont {Bloch},
  \citenamefont {Dalibard},\ and\ \citenamefont {Zwerger}}]{bloch2008many}%
  \BibitemOpen
  \bibfield  {author} {\bibinfo {author} {\bibfnamefont {I.}~\bibnamefont
  {Bloch}}, \bibinfo {author} {\bibfnamefont {J.}~\bibnamefont {Dalibard}}, \
  and\ \bibinfo {author} {\bibfnamefont {W.}~\bibnamefont {Zwerger}},\ }\href
  {\doibase 10.1103/RevModPhys.80.885} {\bibfield  {journal} {\bibinfo
  {journal} {Reviews of Modern Physics}\ }\textbf {\bibinfo {volume} {80}},\
  \bibinfo {pages} {885} (\bibinfo {year} {2008})}\BibitemShut {NoStop}%
\bibitem [{\citenamefont {Lahaye}\ \emph {et~al.}(2009)\citenamefont {Lahaye},
  \citenamefont {Menotti}, \citenamefont {Santos}, \citenamefont {Lewenstein},\
  and\ \citenamefont {Pfau}}]{lahaye2009physics}%
  \BibitemOpen
  \bibfield  {author} {\bibinfo {author} {\bibfnamefont {T.}~\bibnamefont
  {Lahaye}}, \bibinfo {author} {\bibfnamefont {C.}~\bibnamefont {Menotti}},
  \bibinfo {author} {\bibfnamefont {L.}~\bibnamefont {Santos}}, \bibinfo
  {author} {\bibfnamefont {M.}~\bibnamefont {Lewenstein}}, \ and\ \bibinfo
  {author} {\bibfnamefont {T.}~\bibnamefont {Pfau}},\ }\href {\doibase
  10.1088/0034-4885/72/12/126401} {\bibfield  {journal} {\bibinfo  {journal}
  {Reports on Progress in Physics}\ }\textbf {\bibinfo {volume} {72}},\
  \bibinfo {pages} {126401} (\bibinfo {year} {2009})}\BibitemShut {NoStop}%
\bibitem [{\citenamefont {Georgescu}\ \emph {et~al.}(2014)\citenamefont
  {Georgescu}, \citenamefont {Ashhab},\ and\ \citenamefont
  {Nori}}]{georgescu2014quantum}%
  \BibitemOpen
  \bibfield  {author} {\bibinfo {author} {\bibfnamefont {I.~M.}\ \bibnamefont
  {Georgescu}}, \bibinfo {author} {\bibfnamefont {S.}~\bibnamefont {Ashhab}}, \
  and\ \bibinfo {author} {\bibfnamefont {F.}~\bibnamefont {Nori}},\ }\href
  {\doibase 10.1103/RevModPhys.86.153} {\bibfield  {journal} {\bibinfo
  {journal} {Reviews of Modern Physics}\ }\textbf {\bibinfo {volume} {86}},\
  \bibinfo {pages} {153} (\bibinfo {year} {2014})}\BibitemShut {NoStop}%
\bibitem [{\citenamefont {Kruk}\ \emph {et~al.}(2017)\citenamefont {Kruk},
  \citenamefont {Slobozhanyuk}, \citenamefont {Denkova}, \citenamefont
  {Poddubny}, \citenamefont {Kravchenko}, \citenamefont {Miroshnichenko},
  \citenamefont {Neshev},\ and\ \citenamefont {Kivshar}}]{kruk2017edge}%
  \BibitemOpen
  \bibfield  {author} {\bibinfo {author} {\bibfnamefont {S.}~\bibnamefont
  {Kruk}}, \bibinfo {author} {\bibfnamefont {A.}~\bibnamefont {Slobozhanyuk}},
  \bibinfo {author} {\bibfnamefont {D.}~\bibnamefont {Denkova}}, \bibinfo
  {author} {\bibfnamefont {A.}~\bibnamefont {Poddubny}}, \bibinfo {author}
  {\bibfnamefont {I.}~\bibnamefont {Kravchenko}}, \bibinfo {author}
  {\bibfnamefont {A.}~\bibnamefont {Miroshnichenko}}, \bibinfo {author}
  {\bibfnamefont {D.}~\bibnamefont {Neshev}}, \ and\ \bibinfo {author}
  {\bibfnamefont {Y.}~\bibnamefont {Kivshar}},\ }\href {\doibase
  10.1002/smll.201603190} {\bibfield  {journal} {\bibinfo  {journal} {Small}\
  }\textbf {\bibinfo {volume} {13}},\ \bibinfo {pages} {1603190} (\bibinfo
  {year} {2017})}\BibitemShut {NoStop}%
\bibitem [{\citenamefont {St-Jean}\ \emph {et~al.}(2017)\citenamefont
  {St-Jean}, \citenamefont {Goblot}, \citenamefont {Galopin}, \citenamefont
  {Lema{\^\i}tre}, \citenamefont {Ozawa}, \citenamefont {Le~Gratiet},
  \citenamefont {Sagnes}, \citenamefont {Bloch},\ and\ \citenamefont
  {Amo}}]{st2017lasing}%
  \BibitemOpen
  \bibfield  {author} {\bibinfo {author} {\bibfnamefont {P.}~\bibnamefont
  {St-Jean}}, \bibinfo {author} {\bibfnamefont {V.}~\bibnamefont {Goblot}},
  \bibinfo {author} {\bibfnamefont {E.}~\bibnamefont {Galopin}}, \bibinfo
  {author} {\bibfnamefont {A.}~\bibnamefont {Lema{\^\i}tre}}, \bibinfo {author}
  {\bibfnamefont {T.}~\bibnamefont {Ozawa}}, \bibinfo {author} {\bibfnamefont
  {L.}~\bibnamefont {Le~Gratiet}}, \bibinfo {author} {\bibfnamefont
  {I.}~\bibnamefont {Sagnes}}, \bibinfo {author} {\bibfnamefont
  {J.}~\bibnamefont {Bloch}}, \ and\ \bibinfo {author} {\bibfnamefont
  {A.}~\bibnamefont {Amo}},\ }\href {\doibase 10.1038/s41566-017-0006-2}
  {\bibfield  {journal} {\bibinfo  {journal} {Nature Photonics}\ }\textbf
  {\bibinfo {volume} {11}},\ \bibinfo {pages} {651} (\bibinfo {year}
  {2017})}\BibitemShut {NoStop}%
\bibitem [{\citenamefont {Nakajima}\ \emph {et~al.}(2016)\citenamefont
  {Nakajima}, \citenamefont {Tomita}, \citenamefont {Taie}, \citenamefont
  {Ichinose}, \citenamefont {Ozawa}, \citenamefont {Wang}, \citenamefont
  {Troyer},\ and\ \citenamefont {Takahashi}}]{nakajima2016topological}%
  \BibitemOpen
  \bibfield  {author} {\bibinfo {author} {\bibfnamefont {S.}~\bibnamefont
  {Nakajima}}, \bibinfo {author} {\bibfnamefont {T.}~\bibnamefont {Tomita}},
  \bibinfo {author} {\bibfnamefont {S.}~\bibnamefont {Taie}}, \bibinfo {author}
  {\bibfnamefont {T.}~\bibnamefont {Ichinose}}, \bibinfo {author}
  {\bibfnamefont {H.}~\bibnamefont {Ozawa}}, \bibinfo {author} {\bibfnamefont
  {L.}~\bibnamefont {Wang}}, \bibinfo {author} {\bibfnamefont {M.}~\bibnamefont
  {Troyer}}, \ and\ \bibinfo {author} {\bibfnamefont {Y.}~\bibnamefont
  {Takahashi}},\ }\href {\doibase 10.1038/NPHYS3622} {\bibfield  {journal}
  {\bibinfo  {journal} {Nature Physics}\ }\textbf {\bibinfo {volume} {12}},\
  \bibinfo {pages} {296} (\bibinfo {year} {2016})}\BibitemShut {NoStop}%
\bibitem [{\citenamefont {Lohse}\ \emph {et~al.}(2016)\citenamefont {Lohse},
  \citenamefont {Schweizer}, \citenamefont {Zilberberg}, \citenamefont
  {Aidelsburger},\ and\ \citenamefont {Bloch}}]{lohse2016thouless}%
  \BibitemOpen
  \bibfield  {author} {\bibinfo {author} {\bibfnamefont {M.}~\bibnamefont
  {Lohse}}, \bibinfo {author} {\bibfnamefont {C.}~\bibnamefont {Schweizer}},
  \bibinfo {author} {\bibfnamefont {O.}~\bibnamefont {Zilberberg}}, \bibinfo
  {author} {\bibfnamefont {M.}~\bibnamefont {Aidelsburger}}, \ and\ \bibinfo
  {author} {\bibfnamefont {I.}~\bibnamefont {Bloch}},\ }\href {\doibase
  10.1038/nphys3584} {\bibfield  {journal} {\bibinfo  {journal} {Nature
  Physics}\ }\textbf {\bibinfo {volume} {12}},\ \bibinfo {pages} {350}
  (\bibinfo {year} {2016})}\BibitemShut {NoStop}%
\bibitem [{\citenamefont {de~L{\'e}s{\'e}leuc}\ \emph
  {et~al.}(2019)\citenamefont {de~L{\'e}s{\'e}leuc}, \citenamefont {Lienhard},
  \citenamefont {Scholl}, \citenamefont {Barredo}, \citenamefont {Weber},
  \citenamefont {Lang}, \citenamefont {B{\"u}chler}, \citenamefont {Lahaye},\
  and\ \citenamefont {Browaeys}}]{de2019observation}%
  \BibitemOpen
  \bibfield  {author} {\bibinfo {author} {\bibfnamefont {S.}~\bibnamefont
  {de~L{\'e}s{\'e}leuc}}, \bibinfo {author} {\bibfnamefont {V.}~\bibnamefont
  {Lienhard}}, \bibinfo {author} {\bibfnamefont {P.}~\bibnamefont {Scholl}},
  \bibinfo {author} {\bibfnamefont {D.}~\bibnamefont {Barredo}}, \bibinfo
  {author} {\bibfnamefont {S.}~\bibnamefont {Weber}}, \bibinfo {author}
  {\bibfnamefont {N.}~\bibnamefont {Lang}}, \bibinfo {author} {\bibfnamefont
  {H.~P.}\ \bibnamefont {B{\"u}chler}}, \bibinfo {author} {\bibfnamefont
  {T.}~\bibnamefont {Lahaye}}, \ and\ \bibinfo {author} {\bibfnamefont
  {A.}~\bibnamefont {Browaeys}},\ }\href {\doibase 10.1126/science.aav9105}
  {\bibfield  {journal} {\bibinfo  {journal} {Science}\ }\textbf {\bibinfo
  {volume} {365}},\ \bibinfo {pages} {775} (\bibinfo {year}
  {2019})}\BibitemShut {NoStop}%
\bibitem [{\citenamefont {Xie}\ \emph {et~al.}(2019)\citenamefont {Xie},
  \citenamefont {Gou}, \citenamefont {Xiao}, \citenamefont {Gadway},\ and\
  \citenamefont {Yan}}]{xie2019topological}%
  \BibitemOpen
  \bibfield  {author} {\bibinfo {author} {\bibfnamefont {D.}~\bibnamefont
  {Xie}}, \bibinfo {author} {\bibfnamefont {W.}~\bibnamefont {Gou}}, \bibinfo
  {author} {\bibfnamefont {T.}~\bibnamefont {Xiao}}, \bibinfo {author}
  {\bibfnamefont {B.}~\bibnamefont {Gadway}}, \ and\ \bibinfo {author}
  {\bibfnamefont {B.}~\bibnamefont {Yan}},\ }\href {\doibase
  10.1038/s41534-019-0159-6} {\bibfield  {journal} {\bibinfo  {journal} {npj
  Quantum Information}\ }\textbf {\bibinfo {volume} {5}},\ \bibinfo {pages} {1}
  (\bibinfo {year} {2019})}\BibitemShut {NoStop}%
\bibitem [{\citenamefont {Sun}\ \emph {et~al.}(2012)\citenamefont {Sun},
  \citenamefont {Jackeli}, \citenamefont {Santos},\ and\ \citenamefont
  {Vekua}}]{sun2012exploring}%
  \BibitemOpen
  \bibfield  {author} {\bibinfo {author} {\bibfnamefont {G.}~\bibnamefont
  {Sun}}, \bibinfo {author} {\bibfnamefont {G.}~\bibnamefont {Jackeli}},
  \bibinfo {author} {\bibfnamefont {L.}~\bibnamefont {Santos}}, \ and\ \bibinfo
  {author} {\bibfnamefont {T.}~\bibnamefont {Vekua}},\ }\href {\doibase
  10.1103/PhysRevB.86.155159} {\bibfield  {journal} {\bibinfo  {journal}
  {Physical Review B}\ }\textbf {\bibinfo {volume} {86}},\ \bibinfo {pages}
  {155159} (\bibinfo {year} {2012})}\BibitemShut {NoStop}%
\bibitem [{\citenamefont {Pinheiro}\ \emph {et~al.}(2013)\citenamefont
  {Pinheiro}, \citenamefont {Bruun}, \citenamefont {Martikainen},\ and\
  \citenamefont {Larson}}]{pinheiro2013x}%
  \BibitemOpen
  \bibfield  {author} {\bibinfo {author} {\bibfnamefont {F.}~\bibnamefont
  {Pinheiro}}, \bibinfo {author} {\bibfnamefont {G.~M.}\ \bibnamefont {Bruun}},
  \bibinfo {author} {\bibfnamefont {J.-P.}\ \bibnamefont {Martikainen}}, \ and\
  \bibinfo {author} {\bibfnamefont {J.}~\bibnamefont {Larson}},\ }\href
  {\doibase https://doi.org/10.1103/PhysRevLett.111.205302} {\bibfield
  {journal} {\bibinfo  {journal} {Physical review letters}\ }\textbf {\bibinfo
  {volume} {111}},\ \bibinfo {pages} {205302} (\bibinfo {year}
  {2013})}\BibitemShut {NoStop}%
\bibitem [{\citenamefont {Sowi{\'n}ski}\ \emph {et~al.}(2013)\citenamefont
  {Sowi{\'n}ski}, \citenamefont {{\L}{\k{a}}cki}, \citenamefont {Dutta},
  \citenamefont {Pietraszewicz}, \citenamefont {Sierant}, \citenamefont
  {Gajda}, \citenamefont {Zakrzewski},\ and\ \citenamefont
  {Lewenstein}}]{sowinski2013tunneling}%
  \BibitemOpen
  \bibfield  {author} {\bibinfo {author} {\bibfnamefont {T.}~\bibnamefont
  {Sowi{\'n}ski}}, \bibinfo {author} {\bibfnamefont {M.}~\bibnamefont
  {{\L}{\k{a}}cki}}, \bibinfo {author} {\bibfnamefont {O.}~\bibnamefont
  {Dutta}}, \bibinfo {author} {\bibfnamefont {J.}~\bibnamefont
  {Pietraszewicz}}, \bibinfo {author} {\bibfnamefont {P.}~\bibnamefont
  {Sierant}}, \bibinfo {author} {\bibfnamefont {M.}~\bibnamefont {Gajda}},
  \bibinfo {author} {\bibfnamefont {J.}~\bibnamefont {Zakrzewski}}, \ and\
  \bibinfo {author} {\bibfnamefont {M.}~\bibnamefont {Lewenstein}},\ }\href
  {\doibase https://doi.org/10.1103/PhysRevLett.111.215302} {\bibfield
  {journal} {\bibinfo  {journal} {Physical review letters}\ }\textbf {\bibinfo
  {volume} {111}},\ \bibinfo {pages} {215302} (\bibinfo {year}
  {2013})}\BibitemShut {NoStop}%
\bibitem [{\citenamefont {Saugmann}\ and\ \citenamefont
  {Larson}(2020)}]{saugmann2020magnetic}%
  \BibitemOpen
  \bibfield  {author} {\bibinfo {author} {\bibfnamefont {P.}~\bibnamefont
  {Saugmann}}\ and\ \bibinfo {author} {\bibfnamefont {J.}~\bibnamefont
  {Larson}},\ }\href {\doibase 10.1088/1367-2630/ab6cdf} {\bibfield  {journal}
  {\bibinfo  {journal} {New Journal of Physics}\ }\textbf {\bibinfo {volume}
  {22}},\ \bibinfo {pages} {023023} (\bibinfo {year} {2020})}\BibitemShut
  {NoStop}%
\bibitem [{\citenamefont {Isacsson}\ and\ \citenamefont
  {Girvin}(2005)}]{isacsson2005multiflavor}%
  \BibitemOpen
  \bibfield  {author} {\bibinfo {author} {\bibfnamefont {A.}~\bibnamefont
  {Isacsson}}\ and\ \bibinfo {author} {\bibfnamefont {S.~M.}\ \bibnamefont
  {Girvin}},\ }\href {\doibase 10.1103/PhysRevA.72.053604} {\bibfield
  {journal} {\bibinfo  {journal} {Physical Review A}\ }\textbf {\bibinfo
  {volume} {72}},\ \bibinfo {pages} {053604} (\bibinfo {year}
  {2005})}\BibitemShut {NoStop}%
\bibitem [{\citenamefont {M{\"u}ller}\ \emph {et~al.}(2007)\citenamefont
  {M{\"u}ller}, \citenamefont {F{\"o}lling}, \citenamefont {Widera},\ and\
  \citenamefont {Bloch}}]{muller2007state}%
  \BibitemOpen
  \bibfield  {author} {\bibinfo {author} {\bibfnamefont {T.}~\bibnamefont
  {M{\"u}ller}}, \bibinfo {author} {\bibfnamefont {S.}~\bibnamefont
  {F{\"o}lling}}, \bibinfo {author} {\bibfnamefont {A.}~\bibnamefont {Widera}},
  \ and\ \bibinfo {author} {\bibfnamefont {I.}~\bibnamefont {Bloch}},\ }\href
  {\doibase 10.1103/PhysRevLett.99.200405} {\bibfield  {journal} {\bibinfo
  {journal} {Physical Review Letters}\ }\textbf {\bibinfo {volume} {99}},\
  \bibinfo {pages} {200405} (\bibinfo {year} {2007})}\BibitemShut {NoStop}%
\bibitem [{\citenamefont {Wirth}\ \emph {et~al.}(2011)\citenamefont {Wirth},
  \citenamefont {{\"O}lschl{\"a}ger},\ and\ \citenamefont
  {Hemmerich}}]{wirth2011evidence}%
  \BibitemOpen
  \bibfield  {author} {\bibinfo {author} {\bibfnamefont {G.}~\bibnamefont
  {Wirth}}, \bibinfo {author} {\bibfnamefont {M.}~\bibnamefont
  {{\"O}lschl{\"a}ger}}, \ and\ \bibinfo {author} {\bibfnamefont
  {A.}~\bibnamefont {Hemmerich}},\ }\href {\doibase 10.1038/nphys1857}
  {\bibfield  {journal} {\bibinfo  {journal} {Nature Physics}\ }\textbf
  {\bibinfo {volume} {7}},\ \bibinfo {pages} {147} (\bibinfo {year}
  {2011})}\BibitemShut {NoStop}%
\bibitem [{\citenamefont {Niu}\ \emph {et~al.}(2018)\citenamefont {Niu},
  \citenamefont {Jin}, \citenamefont {Chen}, \citenamefont {Li},\ and\
  \citenamefont {Zhou}}]{niu2018observation}%
  \BibitemOpen
  \bibfield  {author} {\bibinfo {author} {\bibfnamefont {L.}~\bibnamefont
  {Niu}}, \bibinfo {author} {\bibfnamefont {S.}~\bibnamefont {Jin}}, \bibinfo
  {author} {\bibfnamefont {X.}~\bibnamefont {Chen}}, \bibinfo {author}
  {\bibfnamefont {X.}~\bibnamefont {Li}}, \ and\ \bibinfo {author}
  {\bibfnamefont {X.}~\bibnamefont {Zhou}},\ }\href {\doibase
  10.1103/PhysRevLett.121.265301} {\bibfield  {journal} {\bibinfo  {journal}
  {Physical Review Letters}\ }\textbf {\bibinfo {volume} {121}},\ \bibinfo
  {pages} {265301} (\bibinfo {year} {2018})}\BibitemShut {NoStop}%
\bibitem [{\citenamefont {Slot}\ \emph {et~al.}(2019)\citenamefont {Slot},
  \citenamefont {Kempkes}, \citenamefont {Knol}, \citenamefont
  {Van~Weerdenburg}, \citenamefont {vandenBroeke}, \citenamefont {Wegner},
  \citenamefont {Vanmaekelbergh}, \citenamefont {Khajetoorians}, \citenamefont
  {MoraisSmith},\ and\ \citenamefont {Swart}}]{slot2019p}%
  \BibitemOpen
  \bibfield  {author} {\bibinfo {author} {\bibfnamefont {M.~R.}\ \bibnamefont
  {Slot}}, \bibinfo {author} {\bibfnamefont {S.~N.}\ \bibnamefont {Kempkes}},
  \bibinfo {author} {\bibfnamefont {E.~J.}\ \bibnamefont {Knol}}, \bibinfo
  {author} {\bibfnamefont {W.~M.~J.}\ \bibnamefont {Van~Weerdenburg}}, \bibinfo
  {author} {\bibfnamefont {J.}~\bibnamefont {vandenBroeke}}, \bibinfo {author}
  {\bibfnamefont {D.}~\bibnamefont {Wegner}}, \bibinfo {author} {\bibfnamefont
  {D.}~\bibnamefont {Vanmaekelbergh}}, \bibinfo {author} {\bibfnamefont
  {A.~A.}\ \bibnamefont {Khajetoorians}}, \bibinfo {author} {\bibfnamefont
  {C.}~\bibnamefont {MoraisSmith}}, \ and\ \bibinfo {author} {\bibfnamefont
  {I.}~\bibnamefont {Swart}},\ }\href {\doibase 10.1103/PhysRevX.9.011009}
  {\bibfield  {journal} {\bibinfo  {journal} {Physical Review X}\ }\textbf
  {\bibinfo {volume} {9}},\ \bibinfo {pages} {011009} (\bibinfo {year}
  {2019})}\BibitemShut {NoStop}%
\bibitem [{\citenamefont {Liu}\ and\ \citenamefont {Wu}(2006)}]{liu2006atomic}%
  \BibitemOpen
  \bibfield  {author} {\bibinfo {author} {\bibfnamefont {W.~V.}\ \bibnamefont
  {Liu}}\ and\ \bibinfo {author} {\bibfnamefont {C.}~\bibnamefont {Wu}},\
  }\href {\doibase 10.1103/PhysRevA.74.013607} {\bibfield  {journal} {\bibinfo
  {journal} {Physical Review A}\ }\textbf {\bibinfo {volume} {74}},\ \bibinfo
  {pages} {013607} (\bibinfo {year} {2006})}\BibitemShut {NoStop}%
\bibitem [{\citenamefont {Wu}\ \emph {et~al.}(2007)\citenamefont {Wu},
  \citenamefont {Bergman}, \citenamefont {Balents},\ and\ \citenamefont
  {DasSarma}}]{wu2007flat}%
  \BibitemOpen
  \bibfield  {author} {\bibinfo {author} {\bibfnamefont {C.}~\bibnamefont
  {Wu}}, \bibinfo {author} {\bibfnamefont {D.}~\bibnamefont {Bergman}},
  \bibinfo {author} {\bibfnamefont {L.}~\bibnamefont {Balents}}, \ and\
  \bibinfo {author} {\bibfnamefont {S.}~\bibnamefont {DasSarma}},\ }\href
  {\doibase 10.1103/PhysRevLett.99.070401} {\bibfield  {journal} {\bibinfo
  {journal} {Physical Review Letters}\ }\textbf {\bibinfo {volume} {99}},\
  \bibinfo {pages} {070401} (\bibinfo {year} {2007})}\BibitemShut {NoStop}%
\bibitem [{\citenamefont {Wu}(2008)}]{wu2008orbital}%
  \BibitemOpen
  \bibfield  {author} {\bibinfo {author} {\bibfnamefont {C.}~\bibnamefont
  {Wu}},\ }\href {\doibase 10.1103/PhysRevLett.101.186807} {\bibfield
  {journal} {\bibinfo  {journal} {Physical Review Letters}\ }\textbf {\bibinfo
  {volume} {101}},\ \bibinfo {pages} {186807} (\bibinfo {year}
  {2008})}\BibitemShut {NoStop}%
\bibitem [{\citenamefont {Zhao}\ and\ \citenamefont
  {Liu}(2008)}]{zhao2008orbital}%
  \BibitemOpen
  \bibfield  {author} {\bibinfo {author} {\bibfnamefont {E.}~\bibnamefont
  {Zhao}}\ and\ \bibinfo {author} {\bibfnamefont {W.~V.}\ \bibnamefont {Liu}},\
  }\href {\doibase 10.1103/PhysRevLett.100.160403} {\bibfield  {journal}
  {\bibinfo  {journal} {Physical Review Letters}\ }\textbf {\bibinfo {volume}
  {100}},\ \bibinfo {pages} {160403} (\bibinfo {year} {2008})}\BibitemShut
  {NoStop}%
\bibitem [{\citenamefont {Lu}\ and\ \citenamefont
  {Arrigoni}(2009)}]{lu2009dispersive}%
  \BibitemOpen
  \bibfield  {author} {\bibinfo {author} {\bibfnamefont {X.}~\bibnamefont
  {Lu}}\ and\ \bibinfo {author} {\bibfnamefont {E.}~\bibnamefont {Arrigoni}},\
  }\href {\doibase 10.1103/PhysRevB.79.245109} {\bibfield  {journal} {\bibinfo
  {journal} {Physical Review B}\ }\textbf {\bibinfo {volume} {79}},\ \bibinfo
  {pages} {245109} (\bibinfo {year} {2009})}\BibitemShut {NoStop}%
\bibitem [{\citenamefont {Wu}(2009)}]{wu2009unconventional}%
  \BibitemOpen
  \bibfield  {author} {\bibinfo {author} {\bibfnamefont {C.}~\bibnamefont
  {Wu}},\ }\href {\doibase 10.1142/S0217984909017777} {\bibfield  {journal}
  {\bibinfo  {journal} {Modern Physics Letters B}\ }\textbf {\bibinfo {volume}
  {23}},\ \bibinfo {pages} {1} (\bibinfo {year} {2009})}\BibitemShut {NoStop}%
\bibitem [{\citenamefont {Hauke}\ \emph {et~al.}(2011)\citenamefont {Hauke},
  \citenamefont {Zhao}, \citenamefont {Goyal}, \citenamefont {Deutsch},
  \citenamefont {Liu},\ and\ \citenamefont {Lewenstein}}]{hauke2011orbital}%
  \BibitemOpen
  \bibfield  {author} {\bibinfo {author} {\bibfnamefont {P.}~\bibnamefont
  {Hauke}}, \bibinfo {author} {\bibfnamefont {E.}~\bibnamefont {Zhao}},
  \bibinfo {author} {\bibfnamefont {K.}~\bibnamefont {Goyal}}, \bibinfo
  {author} {\bibfnamefont {I.~H.}\ \bibnamefont {Deutsch}}, \bibinfo {author}
  {\bibfnamefont {W.~V.}\ \bibnamefont {Liu}}, \ and\ \bibinfo {author}
  {\bibfnamefont {M.}~\bibnamefont {Lewenstein}},\ }\href {\doibase
  10.1103/PhysRevA.84.051603} {\bibfield  {journal} {\bibinfo  {journal}
  {Physical Review A}\ }\textbf {\bibinfo {volume} {84}},\ \bibinfo {pages}
  {051603(R)} (\bibinfo {year} {2011})}\BibitemShut {NoStop}%
\bibitem [{\citenamefont {Kobayashi}\ \emph {et~al.}(2012)\citenamefont
  {Kobayashi}, \citenamefont {Okumura}, \citenamefont {Ota}, \citenamefont
  {Yamada},\ and\ \citenamefont {Machida}}]{kobayashi2012nontrivial}%
  \BibitemOpen
  \bibfield  {author} {\bibinfo {author} {\bibfnamefont {K.}~\bibnamefont
  {Kobayashi}}, \bibinfo {author} {\bibfnamefont {M.}~\bibnamefont {Okumura}},
  \bibinfo {author} {\bibfnamefont {Y.}~\bibnamefont {Ota}}, \bibinfo {author}
  {\bibfnamefont {S.}~\bibnamefont {Yamada}}, \ and\ \bibinfo {author}
  {\bibfnamefont {M.}~\bibnamefont {Machida}},\ }\href {\doibase
  10.1103/PhysRevLett.109.235302} {\bibfield  {journal} {\bibinfo  {journal}
  {Physical Review Letters}\ }\textbf {\bibinfo {volume} {109}},\ \bibinfo
  {pages} {235302} (\bibinfo {year} {2012})}\BibitemShut {NoStop}%
\bibitem [{\citenamefont {Soltan-Panahi}\ \emph {et~al.}(2012)\citenamefont
  {Soltan-Panahi}, \citenamefont {L{\"u}hmann}, \citenamefont {Struck},
  \citenamefont {Windpassinger},\ and\ \citenamefont
  {Sengstock}}]{soltan2012quantum}%
  \BibitemOpen
  \bibfield  {author} {\bibinfo {author} {\bibfnamefont {P.}~\bibnamefont
  {Soltan-Panahi}}, \bibinfo {author} {\bibfnamefont {D.-S.}\ \bibnamefont
  {L{\"u}hmann}}, \bibinfo {author} {\bibfnamefont {J.}~\bibnamefont {Struck}},
  \bibinfo {author} {\bibfnamefont {P.}~\bibnamefont {Windpassinger}}, \ and\
  \bibinfo {author} {\bibfnamefont {K.}~\bibnamefont {Sengstock}},\ }\href
  {\doibase 10.1038/nphys2128} {\bibfield  {journal} {\bibinfo  {journal}
  {Nature Physics}\ }\textbf {\bibinfo {volume} {8}},\ \bibinfo {pages} {71}
  (\bibinfo {year} {2012})}\BibitemShut {NoStop}%
\bibitem [{\citenamefont {Li}\ \emph {et~al.}(2012)\citenamefont {Li},
  \citenamefont {Zhang},\ and\ \citenamefont {Liu}}]{li2012time}%
  \BibitemOpen
  \bibfield  {author} {\bibinfo {author} {\bibfnamefont {X.}~\bibnamefont
  {Li}}, \bibinfo {author} {\bibfnamefont {Z.}~\bibnamefont {Zhang}}, \ and\
  \bibinfo {author} {\bibfnamefont {W.~V.}\ \bibnamefont {Liu}},\ }\href
  {\doibase 10.1103/PhysRevLett.108.175302} {\bibfield  {journal} {\bibinfo
  {journal} {Physical Review Letters}\ }\textbf {\bibinfo {volume} {108}},\
  \bibinfo {pages} {175302} (\bibinfo {year} {2012})}\BibitemShut {NoStop}%
\bibitem [{\citenamefont {Wu}\ \emph {et~al.}(2012)\citenamefont {Wu},
  \citenamefont {He}, \citenamefont {Zang},\ and\ \citenamefont
  {Kou}}]{wu2012topological}%
  \BibitemOpen
  \bibfield  {author} {\bibinfo {author} {\bibfnamefont {Y.-J.}\ \bibnamefont
  {Wu}}, \bibinfo {author} {\bibfnamefont {J.}~\bibnamefont {He}}, \bibinfo
  {author} {\bibfnamefont {C.-L.}\ \bibnamefont {Zang}}, \ and\ \bibinfo
  {author} {\bibfnamefont {S.-P.}\ \bibnamefont {Kou}},\ }\href {\doibase
  10.1103/PhysRevB.86.085128} {\bibfield  {journal} {\bibinfo  {journal}
  {Physical Review B}\ }\textbf {\bibinfo {volume} {86}},\ \bibinfo {pages}
  {085128} (\bibinfo {year} {2012})}\BibitemShut {NoStop}%
\bibitem [{\citenamefont {Li}\ \emph {et~al.}(2013)\citenamefont {Li},
  \citenamefont {Zhao},\ and\ \citenamefont {Liu}}]{li2013topological}%
  \BibitemOpen
  \bibfield  {author} {\bibinfo {author} {\bibfnamefont {X.}~\bibnamefont
  {Li}}, \bibinfo {author} {\bibfnamefont {E.}~\bibnamefont {Zhao}}, \ and\
  \bibinfo {author} {\bibfnamefont {W.~V.}\ \bibnamefont {Liu}},\ }\href
  {\doibase 10.1038/ncomms2523} {\bibfield  {journal} {\bibinfo  {journal}
  {Nature Communications}\ }\textbf {\bibinfo {volume} {4}},\ \bibinfo {pages}
  {1} (\bibinfo {year} {2013})}\BibitemShut {NoStop}%
\bibitem [{\citenamefont {Sun}\ \emph {et~al.}(2014)\citenamefont {Sun},
  \citenamefont {Kolezhuk}, \citenamefont {Santos},\ and\ \citenamefont
  {Vekua}}]{sun2014ferromagnetic}%
  \BibitemOpen
  \bibfield  {author} {\bibinfo {author} {\bibfnamefont {G.}~\bibnamefont
  {Sun}}, \bibinfo {author} {\bibfnamefont {A.~K.}\ \bibnamefont {Kolezhuk}},
  \bibinfo {author} {\bibfnamefont {L.}~\bibnamefont {Santos}}, \ and\ \bibinfo
  {author} {\bibfnamefont {T.}~\bibnamefont {Vekua}},\ }\href {\doibase
  10.1103/PhysRevB.89.134420} {\bibfield  {journal} {\bibinfo  {journal}
  {Physical Review B}\ }\textbf {\bibinfo {volume} {89}},\ \bibinfo {pages}
  {134420} (\bibinfo {year} {2014})}\BibitemShut {NoStop}%
\bibitem [{\citenamefont {You}\ \emph {et~al.}(2014{\natexlab{a}})\citenamefont
  {You}, \citenamefont {Horsch},\ and\ \citenamefont
  {Ole{\'s}}}]{you2014quantum}%
  \BibitemOpen
  \bibfield  {author} {\bibinfo {author} {\bibfnamefont {W.-L.}\ \bibnamefont
  {You}}, \bibinfo {author} {\bibfnamefont {P.}~\bibnamefont {Horsch}}, \ and\
  \bibinfo {author} {\bibfnamefont {A.~M.}\ \bibnamefont {Ole{\'s}}},\ }\href
  {\doibase 10.1103/PhysRevB.89.104425} {\bibfield  {journal} {\bibinfo
  {journal} {Physical Review B}\ }\textbf {\bibinfo {volume} {89}},\ \bibinfo
  {pages} {104425} (\bibinfo {year} {2014}{\natexlab{a}})}\BibitemShut
  {NoStop}%
\bibitem [{\citenamefont {Zhou}\ \emph {et~al.}(2015)\citenamefont {Zhou},
  \citenamefont {Zhao},\ and\ \citenamefont {Liu}}]{zhou2015spin}%
  \BibitemOpen
  \bibfield  {author} {\bibinfo {author} {\bibfnamefont {Z.}~\bibnamefont
  {Zhou}}, \bibinfo {author} {\bibfnamefont {E.}~\bibnamefont {Zhao}}, \ and\
  \bibinfo {author} {\bibfnamefont {W.~V.}\ \bibnamefont {Liu}},\ }\href
  {\doibase 10.1103/PhysRevLett.114.100406} {\bibfield  {journal} {\bibinfo
  {journal} {Physical Review Letters}\ }\textbf {\bibinfo {volume} {114}},\
  \bibinfo {pages} {100406} (\bibinfo {year} {2015})}\BibitemShut {NoStop}%
\bibitem [{\citenamefont {Dutta}\ \emph {et~al.}(2015)\citenamefont {Dutta},
  \citenamefont {Gajda}, \citenamefont {Hauke}, \citenamefont {Lewenstein},
  \citenamefont {L{\"u}hmann}, \citenamefont {Malomed}, \citenamefont
  {Sowi{\'n}ski},\ and\ \citenamefont {Zakrzewski}}]{dutta2015non}%
  \BibitemOpen
  \bibfield  {author} {\bibinfo {author} {\bibfnamefont {O.}~\bibnamefont
  {Dutta}}, \bibinfo {author} {\bibfnamefont {M.}~\bibnamefont {Gajda}},
  \bibinfo {author} {\bibfnamefont {P.}~\bibnamefont {Hauke}}, \bibinfo
  {author} {\bibfnamefont {M.}~\bibnamefont {Lewenstein}}, \bibinfo {author}
  {\bibfnamefont {D.-S.}\ \bibnamefont {L{\"u}hmann}}, \bibinfo {author}
  {\bibfnamefont {B.~A.}\ \bibnamefont {Malomed}}, \bibinfo {author}
  {\bibfnamefont {T.}~\bibnamefont {Sowi{\'n}ski}}, \ and\ \bibinfo {author}
  {\bibfnamefont {J.}~\bibnamefont {Zakrzewski}},\ }\href {\doibase
  10.1088/0034-4885/78/6/066001} {\bibfield  {journal} {\bibinfo  {journal}
  {Reports on Progress in Physics}\ }\textbf {\bibinfo {volume} {78}},\
  \bibinfo {pages} {066001} (\bibinfo {year} {2015})}\BibitemShut {NoStop}%
\bibitem [{\citenamefont {Li}\ and\ \citenamefont {Liu}(2016)}]{li2016physics}%
  \BibitemOpen
  \bibfield  {author} {\bibinfo {author} {\bibfnamefont {X.}~\bibnamefont
  {Li}}\ and\ \bibinfo {author} {\bibfnamefont {W.~V.}\ \bibnamefont {Liu}},\
  }\href {\doibase 10.1088/0034-4885/79/11/116401} {\bibfield  {journal}
  {\bibinfo  {journal} {Reports on Progress in Physics}\ }\textbf {\bibinfo
  {volume} {79}},\ \bibinfo {pages} {116401} (\bibinfo {year}
  {2016})}\BibitemShut {NoStop}%
\bibitem [{\citenamefont {Xu}\ \emph {et~al.}(2016)\citenamefont {Xu},
  \citenamefont {You}, \citenamefont {Hemmerich},\ and\ \citenamefont
  {Liu}}]{xu2016pi}%
  \BibitemOpen
  \bibfield  {author} {\bibinfo {author} {\bibfnamefont {Z.-F.}\ \bibnamefont
  {Xu}}, \bibinfo {author} {\bibfnamefont {L.}~\bibnamefont {You}}, \bibinfo
  {author} {\bibfnamefont {A.}~\bibnamefont {Hemmerich}}, \ and\ \bibinfo
  {author} {\bibfnamefont {W.~V.}\ \bibnamefont {Liu}},\ }\href {\doibase
  10.1103/PhysRevLett.117.085301} {\bibfield  {journal} {\bibinfo  {journal}
  {Physical Review Letters}\ }\textbf {\bibinfo {volume} {117}},\ \bibinfo
  {pages} {085301} (\bibinfo {year} {2016})}\BibitemShut {NoStop}%
\bibitem [{\citenamefont {Liu}\ \emph {et~al.}(2018)\citenamefont {Liu},
  \citenamefont {Zhang}, \citenamefont {Gao},\ and\ \citenamefont
  {Li}}]{liu2018chiral}%
  \BibitemOpen
  \bibfield  {author} {\bibinfo {author} {\bibfnamefont {B.}~\bibnamefont
  {Liu}}, \bibinfo {author} {\bibfnamefont {P.}~\bibnamefont {Zhang}}, \bibinfo
  {author} {\bibfnamefont {H.}~\bibnamefont {Gao}}, \ and\ \bibinfo {author}
  {\bibfnamefont {F.}~\bibnamefont {Li}},\ }\href {\doibase
  10.1103/PhysRevLett.121.015303} {\bibfield  {journal} {\bibinfo  {journal}
  {Physical Review Letters}\ }\textbf {\bibinfo {volume} {121}},\ \bibinfo
  {pages} {015303} (\bibinfo {year} {2018})}\BibitemShut {NoStop}%
\bibitem [{\citenamefont {Li}\ \emph {et~al.}(2018)\citenamefont {Li},
  \citenamefont {Yuan}, \citenamefont {Hemmerich},\ and\ \citenamefont
  {Li}}]{li2018rotation}%
  \BibitemOpen
  \bibfield  {author} {\bibinfo {author} {\bibfnamefont {Y.}~\bibnamefont
  {Li}}, \bibinfo {author} {\bibfnamefont {J.}~\bibnamefont {Yuan}}, \bibinfo
  {author} {\bibfnamefont {A.}~\bibnamefont {Hemmerich}}, \ and\ \bibinfo
  {author} {\bibfnamefont {X.}~\bibnamefont {Li}},\ }\href {\doibase
  10.1103/PhysRevLett.121.093401} {\bibfield  {journal} {\bibinfo  {journal}
  {Physical Review Letters}\ }\textbf {\bibinfo {volume} {121}},\ \bibinfo
  {pages} {093401} (\bibinfo {year} {2018})}\BibitemShut {NoStop}%
\bibitem [{\citenamefont {Jin}\ \emph {et~al.}(2019)\citenamefont {Jin},
  \citenamefont {Zhang}, \citenamefont {Guo}, \citenamefont {Chen},
  \citenamefont {Zhou},\ and\ \citenamefont {Li}}]{jin2019dynamical}%
  \BibitemOpen
  \bibfield  {author} {\bibinfo {author} {\bibfnamefont {S.}~\bibnamefont
  {Jin}}, \bibinfo {author} {\bibfnamefont {W.}~\bibnamefont {Zhang}}, \bibinfo
  {author} {\bibfnamefont {X.}~\bibnamefont {Guo}}, \bibinfo {author}
  {\bibfnamefont {X.}~\bibnamefont {Chen}}, \bibinfo {author} {\bibfnamefont
  {X.}~\bibnamefont {Zhou}}, \ and\ \bibinfo {author} {\bibfnamefont
  {X.}~\bibnamefont {Li}},\ }\href {https://arxiv.org/abs/1910.11880}
  {\bibfield  {journal} {\bibinfo  {journal} {arXiv preprint arXiv:1910.11880}\
  } (\bibinfo {year} {2019})}\BibitemShut {NoStop}%
\bibitem [{\citenamefont {Zhu}\ \emph {et~al.}(2019)\citenamefont {Zhu},
  \citenamefont {Sun}, \citenamefont {Yang}, \citenamefont {Wang},
  \citenamefont {Liu},\ and\ \citenamefont {Ji}}]{zhu2019interaction}%
  \BibitemOpen
  \bibfield  {author} {\bibinfo {author} {\bibfnamefont {G.-B.}\ \bibnamefont
  {Zhu}}, \bibinfo {author} {\bibfnamefont {Q.}~\bibnamefont {Sun}}, \bibinfo
  {author} {\bibfnamefont {H.-M.}\ \bibnamefont {Yang}}, \bibinfo {author}
  {\bibfnamefont {L.-L.}\ \bibnamefont {Wang}}, \bibinfo {author}
  {\bibfnamefont {W.-M.}\ \bibnamefont {Liu}}, \ and\ \bibinfo {author}
  {\bibfnamefont {A.-C.}\ \bibnamefont {Ji}},\ }\href {\doibase
  10.1103/PhysRevA.100.043608} {\bibfield  {journal} {\bibinfo  {journal}
  {Physical Review A}\ }\textbf {\bibinfo {volume} {100}},\ \bibinfo {pages}
  {043608} (\bibinfo {year} {2019})}\BibitemShut {NoStop}%
\bibitem [{\citenamefont {Lee}(2016)}]{lee2016anomalous}%
  \BibitemOpen
  \bibfield  {author} {\bibinfo {author} {\bibfnamefont {T.~E.}\ \bibnamefont
  {Lee}},\ }\href {\doibase 10.1103/PhysRevLett.116.133903} {\bibfield
  {journal} {\bibinfo  {journal} {Physical review letters}\ }\textbf {\bibinfo
  {volume} {116}},\ \bibinfo {pages} {133903} (\bibinfo {year}
  {2016})}\BibitemShut {NoStop}%
\bibitem [{\citenamefont {Yao}\ and\ \citenamefont {Wang}(2018)}]{yao2018edge}%
  \BibitemOpen
  \bibfield  {author} {\bibinfo {author} {\bibfnamefont {S.}~\bibnamefont
  {Yao}}\ and\ \bibinfo {author} {\bibfnamefont {Z.}~\bibnamefont {Wang}},\
  }\href {\doibase 10.1103/PhysRevLett.121.086803} {\bibfield  {journal}
  {\bibinfo  {journal} {Physical review letters}\ }\textbf {\bibinfo {volume}
  {121}},\ \bibinfo {pages} {086803} (\bibinfo {year} {2018})}\BibitemShut
  {NoStop}%
\bibitem [{\citenamefont {Wohlfeld}\ \emph {et~al.}(2008)\citenamefont
  {Wohlfeld}, \citenamefont {Daghofer}, \citenamefont
  {Ole\ifmmode~\acute{s}\else \'{s}\fi{}},\ and\ \citenamefont
  {Horsch}}]{PhysRevB.78.214423}%
  \BibitemOpen
  \bibfield  {author} {\bibinfo {author} {\bibfnamefont {K.}~\bibnamefont
  {Wohlfeld}}, \bibinfo {author} {\bibfnamefont {M.}~\bibnamefont {Daghofer}},
  \bibinfo {author} {\bibfnamefont {A.~M.}\ \bibnamefont
  {Ole\ifmmode~\acute{s}\else \'{s}\fi{}}}, \ and\ \bibinfo {author}
  {\bibfnamefont {P.}~\bibnamefont {Horsch}},\ }\href {\doibase
  10.1103/PhysRevB.78.214423} {\bibfield  {journal} {\bibinfo  {journal}
  {Physical Review B}\ }\textbf {\bibinfo {volume} {78}},\ \bibinfo {pages}
  {214423} (\bibinfo {year} {2008})}\BibitemShut {NoStop}%
\bibitem [{\citenamefont {Daghofer}\ \emph {et~al.}(2008)\citenamefont
  {Daghofer}, \citenamefont {Wohlfeld}, \citenamefont
  {Ole\ifmmode~\acute{s}\else \'{s}\fi{}}, \citenamefont {Arrigoni},\ and\
  \citenamefont {Horsch}}]{Daghofer2008absence}%
  \BibitemOpen
  \bibfield  {author} {\bibinfo {author} {\bibfnamefont {M.}~\bibnamefont
  {Daghofer}}, \bibinfo {author} {\bibfnamefont {K.}~\bibnamefont {Wohlfeld}},
  \bibinfo {author} {\bibfnamefont {A.~M.}\ \bibnamefont
  {Ole\ifmmode~\acute{s}\else \'{s}\fi{}}}, \bibinfo {author} {\bibfnamefont
  {E.}~\bibnamefont {Arrigoni}}, \ and\ \bibinfo {author} {\bibfnamefont
  {P.}~\bibnamefont {Horsch}},\ }\href {\doibase
  10.1103/PhysRevLett.100.066403} {\bibfield  {journal} {\bibinfo  {journal}
  {Physical review letters}\ }\textbf {\bibinfo {volume} {100}},\ \bibinfo
  {pages} {066403} (\bibinfo {year} {2008})}\BibitemShut {NoStop}%
\bibitem [{\citenamefont {You}\ \emph {et~al.}(2007)\citenamefont {You},
  \citenamefont {Li},\ and\ \citenamefont {Gu}}]{PhysRevE.76.022101}%
  \BibitemOpen
  \bibfield  {author} {\bibinfo {author} {\bibfnamefont {W.-L.}\ \bibnamefont
  {You}}, \bibinfo {author} {\bibfnamefont {Y.-W.}\ \bibnamefont {Li}}, \ and\
  \bibinfo {author} {\bibfnamefont {S.-J.}\ \bibnamefont {Gu}},\ }\href
  {\doibase 10.1103/PhysRevE.76.022101} {\bibfield  {journal} {\bibinfo
  {journal} {Physical Review E}\ }\textbf {\bibinfo {volume} {76}},\ \bibinfo
  {pages} {022101} (\bibinfo {year} {2007})}\BibitemShut {NoStop}%
\bibitem [{\citenamefont {Gu}(2010)}]{gu2010fidelity}%
  \BibitemOpen
  \bibfield  {author} {\bibinfo {author} {\bibfnamefont {S.-J.}\ \bibnamefont
  {Gu}},\ }\href {\doibase 10.1142/S0217979210056335} {\bibfield  {journal}
  {\bibinfo  {journal} {International Journal of Modern Physics B}\ }\textbf
  {\bibinfo {volume} {24}},\ \bibinfo {pages} {4371} (\bibinfo {year}
  {2010})}\BibitemShut {NoStop}%
\bibitem [{\citenamefont {Sun}(2017)}]{sun2017fidelity}%
  \BibitemOpen
  \bibfield  {author} {\bibinfo {author} {\bibfnamefont {G.}~\bibnamefont
  {Sun}},\ }\href {\doibase 10.1103/PhysRevA.96.043621} {\bibfield  {journal}
  {\bibinfo  {journal} {Physical Review A}\ }\textbf {\bibinfo {volume} {96}},\
  \bibinfo {pages} {043621} (\bibinfo {year} {2017})}\BibitemShut {NoStop}%
\bibitem [{\citenamefont {Zhu}\ \emph {et~al.}(2018)\citenamefont {Zhu},
  \citenamefont {Sun}, \citenamefont {You},\ and\ \citenamefont
  {Shi}}]{zhu2018fidelity}%
  \BibitemOpen
  \bibfield  {author} {\bibinfo {author} {\bibfnamefont {Z.}~\bibnamefont
  {Zhu}}, \bibinfo {author} {\bibfnamefont {G.}~\bibnamefont {Sun}}, \bibinfo
  {author} {\bibfnamefont {W.-L.}\ \bibnamefont {You}}, \ and\ \bibinfo
  {author} {\bibfnamefont {D.-N.}\ \bibnamefont {Shi}},\ }\href {\doibase
  10.1103/PhysRevA.98.023607} {\bibfield  {journal} {\bibinfo  {journal}
  {Physical Review A}\ }\textbf {\bibinfo {volume} {98}},\ \bibinfo {pages}
  {023607} (\bibinfo {year} {2018})}\BibitemShut {NoStop}%
\bibitem [{\citenamefont {Ren}\ \emph {et~al.}(2020{\natexlab{a}})\citenamefont
  {Ren}, \citenamefont {You},\ and\ \citenamefont
  {Wang}}]{PhysRevB.101.094410}%
  \BibitemOpen
  \bibfield  {author} {\bibinfo {author} {\bibfnamefont {J.}~\bibnamefont
  {Ren}}, \bibinfo {author} {\bibfnamefont {W.-L.}\ \bibnamefont {You}}, \ and\
  \bibinfo {author} {\bibfnamefont {X.}~\bibnamefont {Wang}},\ }\href {\doibase
  10.1103/PhysRevB.101.094410} {\bibfield  {journal} {\bibinfo  {journal}
  {Physical Review B}\ }\textbf {\bibinfo {volume} {101}},\ \bibinfo {pages}
  {094410} (\bibinfo {year} {2020}{\natexlab{a}})}\BibitemShut {NoStop}%
\bibitem [{\citenamefont {Ren}\ \emph {et~al.}(2020{\natexlab{b}})\citenamefont
  {Ren}, \citenamefont {You},\ and\ \citenamefont {Ole\ifmmode~\acute{s}\else
  \'{s}\fi{}}}]{PhysRevB.102.024425}%
  \BibitemOpen
  \bibfield  {author} {\bibinfo {author} {\bibfnamefont {J.}~\bibnamefont
  {Ren}}, \bibinfo {author} {\bibfnamefont {W.-L.}\ \bibnamefont {You}}, \ and\
  \bibinfo {author} {\bibfnamefont {A.~M.}\ \bibnamefont
  {Ole\ifmmode~\acute{s}\else \'{s}\fi{}}},\ }\href {\doibase
  10.1103/PhysRevB.102.024425} {\bibfield  {journal} {\bibinfo  {journal}
  {Physical Review B}\ }\textbf {\bibinfo {volume} {102}},\ \bibinfo {pages}
  {024425} (\bibinfo {year} {2020}{\natexlab{b}})}\BibitemShut {NoStop}%
\bibitem [{\citenamefont {Heyl}\ \emph {et~al.}(2013)\citenamefont {Heyl},
  \citenamefont {Polkovnikov},\ and\ \citenamefont
  {Kehrein}}]{heyl2013dynamical}%
  \BibitemOpen
  \bibfield  {author} {\bibinfo {author} {\bibfnamefont {M.}~\bibnamefont
  {Heyl}}, \bibinfo {author} {\bibfnamefont {A.}~\bibnamefont {Polkovnikov}}, \
  and\ \bibinfo {author} {\bibfnamefont {S.}~\bibnamefont {Kehrein}},\ }\href
  {\doibase 10.1103/PhysRevLett.110.135704} {\bibfield  {journal} {\bibinfo
  {journal} {Physical review letters}\ }\textbf {\bibinfo {volume} {110}},\
  \bibinfo {pages} {135704} (\bibinfo {year} {2013})}\BibitemShut {NoStop}%
\bibitem [{\citenamefont {Heyl}(2018)}]{heyl2018dynamical}%
  \BibitemOpen
  \bibfield  {author} {\bibinfo {author} {\bibfnamefont {M.}~\bibnamefont
  {Heyl}},\ }\href {\doibase 10.1088/1361-6633/aaaf9a} {\bibfield  {journal}
  {\bibinfo  {journal} {Reports on Progress in Physics}\ }\textbf {\bibinfo
  {volume} {81}},\ \bibinfo {pages} {054001} (\bibinfo {year}
  {2018})}\BibitemShut {NoStop}%
\bibitem [{\citenamefont {J{\"u}nemann}\ \emph {et~al.}(2017)\citenamefont
  {J{\"u}nemann}, \citenamefont {Piga}, \citenamefont {Ran}, \citenamefont
  {Lewenstein}, \citenamefont {Rizzi},\ and\ \citenamefont
  {Berm{\'u}dez}}]{junemann2017exploring}%
  \BibitemOpen
  \bibfield  {author} {\bibinfo {author} {\bibfnamefont {J.}~\bibnamefont
  {J{\"u}nemann}}, \bibinfo {author} {\bibfnamefont {A.}~\bibnamefont {Piga}},
  \bibinfo {author} {\bibfnamefont {S.-J.}\ \bibnamefont {Ran}}, \bibinfo
  {author} {\bibfnamefont {M.}~\bibnamefont {Lewenstein}}, \bibinfo {author}
  {\bibfnamefont {M.}~\bibnamefont {Rizzi}}, \ and\ \bibinfo {author}
  {\bibfnamefont {A.}~\bibnamefont {Berm{\'u}dez}},\ }\href {\doibase
  https://doi.org/10.1103/PhysRevX.7.031057} {\bibfield  {journal} {\bibinfo
  {journal} {Physical Review X}\ }\textbf {\bibinfo {volume} {7}},\ \bibinfo
  {pages} {031057} (\bibinfo {year} {2017})}\BibitemShut {NoStop}%
\bibitem [{\citenamefont {Bermudez}\ \emph {et~al.}(2018)\citenamefont
  {Bermudez}, \citenamefont {Tirrito}, \citenamefont {Rizzi}, \citenamefont
  {Lewenstein},\ and\ \citenamefont {Hands}}]{bermudez2018gross}%
  \BibitemOpen
  \bibfield  {author} {\bibinfo {author} {\bibfnamefont {A.}~\bibnamefont
  {Bermudez}}, \bibinfo {author} {\bibfnamefont {E.}~\bibnamefont {Tirrito}},
  \bibinfo {author} {\bibfnamefont {M.}~\bibnamefont {Rizzi}}, \bibinfo
  {author} {\bibfnamefont {M.}~\bibnamefont {Lewenstein}}, \ and\ \bibinfo
  {author} {\bibfnamefont {S.}~\bibnamefont {Hands}},\ }\href {\doibase
  10.1016/j.aop.2018.10.007} {\bibfield  {journal} {\bibinfo  {journal} {Annals
  of Physics}\ }\textbf {\bibinfo {volume} {399}},\ \bibinfo {pages} {149}
  (\bibinfo {year} {2018})}\BibitemShut {NoStop}%
\bibitem [{\citenamefont {Kuno}(2019)}]{kuno2019phase}%
  \BibitemOpen
  \bibfield  {author} {\bibinfo {author} {\bibfnamefont {Y.}~\bibnamefont
  {Kuno}},\ }\href {\doibase 10.1103/PhysRevB.99.064105} {\bibfield  {journal}
  {\bibinfo  {journal} {Physical Review B}\ }\textbf {\bibinfo {volume} {99}},\
  \bibinfo {pages} {064105} (\bibinfo {year} {2019})}\BibitemShut {NoStop}%
\bibitem [{\citenamefont {Kuno}\ and\ \citenamefont
  {Hatsugai}(2020)}]{kuno2020interaction}%
  \BibitemOpen
  \bibfield  {author} {\bibinfo {author} {\bibfnamefont {Y.}~\bibnamefont
  {Kuno}}\ and\ \bibinfo {author} {\bibfnamefont {Y.}~\bibnamefont
  {Hatsugai}},\ }\href {https://arxiv.org/abs/2007.11215} {\bibfield  {journal}
  {\bibinfo  {journal} {arXiv e-prints}\ ,\ \bibinfo {pages} {arXiv}} (\bibinfo
  {year} {2020})}\BibitemShut {NoStop}%
\bibitem [{\citenamefont {Zhu}\ \emph {et~al.}(2013)\citenamefont {Zhu},
  \citenamefont {Wang}, \citenamefont {Chan},\ and\ \citenamefont
  {Duan}}]{zhu2013topological}%
  \BibitemOpen
  \bibfield  {author} {\bibinfo {author} {\bibfnamefont {S.-L.}\ \bibnamefont
  {Zhu}}, \bibinfo {author} {\bibfnamefont {Z.-D.}\ \bibnamefont {Wang}},
  \bibinfo {author} {\bibfnamefont {Y.-H.}\ \bibnamefont {Chan}}, \ and\
  \bibinfo {author} {\bibfnamefont {L.-M.}\ \bibnamefont {Duan}},\ }\href
  {\doibase https://doi.org/10.1103/PhysRevLett.110.075303} {\bibfield
  {journal} {\bibinfo  {journal} {Physical review letters}\ }\textbf {\bibinfo
  {volume} {110}},\ \bibinfo {pages} {075303} (\bibinfo {year}
  {2013})}\BibitemShut {NoStop}%
\bibitem [{\citenamefont {Deng}\ and\ \citenamefont
  {Santos}(2014)}]{deng2014topological}%
  \BibitemOpen
  \bibfield  {author} {\bibinfo {author} {\bibfnamefont {X.}~\bibnamefont
  {Deng}}\ and\ \bibinfo {author} {\bibfnamefont {L.}~\bibnamefont {Santos}},\
  }\href {\doibase https://doi.org/10.1103/PhysRevA.89.033632} {\bibfield
  {journal} {\bibinfo  {journal} {Physical review A}\ }\textbf {\bibinfo
  {volume} {89}},\ \bibinfo {pages} {033632} (\bibinfo {year}
  {2014})}\BibitemShut {NoStop}%
\bibitem [{\citenamefont {Sun}(2016)}]{sun2016topological}%
  \BibitemOpen
  \bibfield  {author} {\bibinfo {author} {\bibfnamefont {G.}~\bibnamefont
  {Sun}},\ }\href {\doibase https://doi.org/10.1103/PhysRevA.93.023608}
  {\bibfield  {journal} {\bibinfo  {journal} {Physical Review A}\ }\textbf
  {\bibinfo {volume} {93}},\ \bibinfo {pages} {023608} (\bibinfo {year}
  {2016})}\BibitemShut {NoStop}%
\bibitem [{\citenamefont {Zhang}\ \emph {et~al.}(2020)\citenamefont {Zhang},
  \citenamefont {Chen}, \citenamefont {Zhang}, \citenamefont {Lang},
  \citenamefont {Li},\ and\ \citenamefont {Zhu}}]{zhang2020skin}%
  \BibitemOpen
  \bibfield  {author} {\bibinfo {author} {\bibfnamefont {D.-W.}\ \bibnamefont
  {Zhang}}, \bibinfo {author} {\bibfnamefont {Y.-L.}\ \bibnamefont {Chen}},
  \bibinfo {author} {\bibfnamefont {G.-Q.}\ \bibnamefont {Zhang}}, \bibinfo
  {author} {\bibfnamefont {L.-J.}\ \bibnamefont {Lang}}, \bibinfo {author}
  {\bibfnamefont {Z.}~\bibnamefont {Li}}, \ and\ \bibinfo {author}
  {\bibfnamefont {S.-L.}\ \bibnamefont {Zhu}},\ }\href {\doibase
  https://doi.org/10.1103/PhysRevB.101.235150} {\bibfield  {journal} {\bibinfo
  {journal} {Physical review B}\ }\textbf {\bibinfo {volume} {101}},\ \bibinfo
  {pages} {235150} (\bibinfo {year} {2020})}\BibitemShut {NoStop}%
\bibitem [{\citenamefont {Sun}\ and\ \citenamefont
  {Wei}(2020)}]{sun2020dynamical}%
  \BibitemOpen
  \bibfield  {author} {\bibinfo {author} {\bibfnamefont {G.}~\bibnamefont
  {Sun}}\ and\ \bibinfo {author} {\bibfnamefont {B.-B.}\ \bibnamefont {Wei}},\
  }\href {https://arxiv.org/abs/2006.00726} {\bibfield  {journal} {\bibinfo
  {journal} {arXiv preprint arXiv:2006.00726}\ } (\bibinfo {year}
  {2020})}\BibitemShut {NoStop}%
\bibitem [{\citenamefont {Ashida}\ \emph {et~al.}(2020)\citenamefont {Ashida},
  \citenamefont {Gong},\ and\ \citenamefont {Ueda}}]{ashida2020non}%
  \BibitemOpen
  \bibfield  {author} {\bibinfo {author} {\bibfnamefont {Y.}~\bibnamefont
  {Ashida}}, \bibinfo {author} {\bibfnamefont {Z.}~\bibnamefont {Gong}}, \ and\
  \bibinfo {author} {\bibfnamefont {M.}~\bibnamefont {Ueda}},\ }\href
  {https://arxiv.org/abs/2006.01837} {\bibfield  {journal} {\bibinfo  {journal}
  {arXiv preprint arXiv:2006.01837}\ } (\bibinfo {year} {2020})}\BibitemShut
  {NoStop}%
\bibitem [{\citenamefont {Pickup}\ \emph {et~al.}(2020)\citenamefont {Pickup},
  \citenamefont {Sigurdsson}, \citenamefont {Ruostekoski},\ and\ \citenamefont
  {Lagoudakis}}]{pickup2020synthetic}%
  \BibitemOpen
  \bibfield  {author} {\bibinfo {author} {\bibfnamefont {L.}~\bibnamefont
  {Pickup}}, \bibinfo {author} {\bibfnamefont {H.}~\bibnamefont {Sigurdsson}},
  \bibinfo {author} {\bibfnamefont {J.}~\bibnamefont {Ruostekoski}}, \ and\
  \bibinfo {author} {\bibfnamefont {P.}~\bibnamefont {Lagoudakis}},\ }\href
  {https://arxiv.org/abs/2001.07616} {\bibfield  {journal} {\bibinfo  {journal}
  {arXiv preprint arXiv:2001.07616}\ } (\bibinfo {year} {2020})}\BibitemShut
  {NoStop}%
\bibitem [{\citenamefont {Jackeli}\ and\ \citenamefont
  {Khaliullin}(2009)}]{PhysRevLett.102.017205}%
  \BibitemOpen
  \bibfield  {author} {\bibinfo {author} {\bibfnamefont {G.}~\bibnamefont
  {Jackeli}}\ and\ \bibinfo {author} {\bibfnamefont {G.}~\bibnamefont
  {Khaliullin}},\ }\href {\doibase 10.1103/PhysRevLett.102.017205} {\bibfield
  {journal} {\bibinfo  {journal} {Physical review letters}\ }\textbf {\bibinfo
  {volume} {102}},\ \bibinfo {pages} {017205} (\bibinfo {year}
  {2009})}\BibitemShut {NoStop}%
\bibitem [{\citenamefont {You}\ \emph {et~al.}(2014{\natexlab{b}})\citenamefont
  {You}, \citenamefont {Horsch},\ and\ \citenamefont
  {Ole\ifmmode~\acute{s}\else \'{s}\fi{}}}]{PhysRevB.89.104425}%
  \BibitemOpen
  \bibfield  {author} {\bibinfo {author} {\bibfnamefont {W.-L.}\ \bibnamefont
  {You}}, \bibinfo {author} {\bibfnamefont {P.}~\bibnamefont {Horsch}}, \ and\
  \bibinfo {author} {\bibfnamefont {A.~M.}\ \bibnamefont
  {Ole\ifmmode~\acute{s}\else \'{s}\fi{}}},\ }\href {\doibase
  10.1103/PhysRevB.89.104425} {\bibfield  {journal} {\bibinfo  {journal}
  {Physical Review B}\ }\textbf {\bibinfo {volume} {89}},\ \bibinfo {pages}
  {104425} (\bibinfo {year} {2014}{\natexlab{b}})}\BibitemShut {NoStop}%
\bibitem [{\citenamefont {Cincio}\ \emph {et~al.}(2010)\citenamefont {Cincio},
  \citenamefont {Dziarmaga},\ and\ \citenamefont {Ole\ifmmode~\acute{s}\else
  \'{s}\fi{}}}]{PhysRevB.82.104416}%
  \BibitemOpen
  \bibfield  {author} {\bibinfo {author} {\bibfnamefont {L.}~\bibnamefont
  {Cincio}}, \bibinfo {author} {\bibfnamefont {J.}~\bibnamefont {Dziarmaga}}, \
  and\ \bibinfo {author} {\bibfnamefont {A.~M.}\ \bibnamefont
  {Ole\ifmmode~\acute{s}\else \'{s}\fi{}}},\ }\href {\doibase
  10.1103/PhysRevB.82.104416} {\bibfield  {journal} {\bibinfo  {journal}
  {Physical Review B}\ }\textbf {\bibinfo {volume} {82}},\ \bibinfo {pages}
  {104416} (\bibinfo {year} {2010})}\BibitemShut {NoStop}%
\bibitem [{\citenamefont {Emery}\ \emph {et~al.}(1990)\citenamefont {Emery},
  \citenamefont {Kivelson},\ and\ \citenamefont {Lin}}]{emery1990phase}%
  \BibitemOpen
  \bibfield  {author} {\bibinfo {author} {\bibfnamefont {V.~J.}\ \bibnamefont
  {Emery}}, \bibinfo {author} {\bibfnamefont {S.~A.}\ \bibnamefont {Kivelson}},
  \ and\ \bibinfo {author} {\bibfnamefont {H.~Q.}\ \bibnamefont {Lin}},\ }\href
  {\doibase 10.1103/PhysRevLett.64.475} {\bibfield  {journal} {\bibinfo
  {journal} {Physical review letters}\ }\textbf {\bibinfo {volume} {64}},\
  \bibinfo {pages} {475} (\bibinfo {year} {1990})}\BibitemShut {NoStop}%
\end{thebibliography}%

\end{document}